\documentclass[%
 reprint,
superscriptaddress,
amsmath,amssymb,
 aps,
]{revtex4-2}

\usepackage[utf8]{inputenc}
\usepackage{graphicx, amsmath}
\usepackage{microtype}
\usepackage{dcolumn}
\usepackage{xcolor,soul}
\usepackage{array}
\usepackage{natbib}
\usepackage{siunitx}
\usepackage{bm}

\usepackage[normalem]{ulem}

\definecolor{colreva}{rgb}{0,0,0.9}
\definecolor{colrevb}{rgb}{0.9,0,0}
\newcommand{\reva}[1]{\textcolor{black}{#1}}
\newcommand{\revb}[1]{\textcolor{black}{#1}}
\begin{document}
\title{A computational model of twisted elastic ribbons}
\author{Madelyn Leembruggen}
\email{mleembruggen@g.harvard.edu}
\affiliation{Department of Physics, Harvard University, Cambridge, MA 02138, USA}
\author{Jovana Andrejevic}
\email{jovana@sas.upenn.edu}
\affiliation{Department of Physics, University of Pennsylvania, Philadelphia, Pennsylvania 19104, USA}
\author{Arshad Kudrolli}
\email{akudrolli@clarku.edu}
\affiliation{Department of Physics, Clark University, Worcester, Massachusetts 01610, USA}
\author{Chris H. Rycroft}
\email{chr@math.wisc.edu}
\affiliation{Department of Mathematics, University of Wisconsin--Madison, Madison, WI 53706, USA}
\affiliation{Computational Research Division, Lawrence Berkeley Laboratory, Berkeley, CA 94720, USA}

\begin{abstract}
We develop an irregular lattice mass-spring-model (MSM) to simulate and study the deformation modes of a thin elastic ribbon as a function of applied end-to-end twist and tension. Our simulations reproduce all reported experimentally observed modes, including transitions from helicoids to longitudinal wrinkles, creased helicoids and loops with self-contact, and transverse wrinkles to accordion self-folds. Our simulations also show that the twist angles at which the primary longitudinal and transverse wrinkles appear are well described by various analyses of the F\"oppl-von K\'arm\'an (FvK) equations, but the characteristic wavelength of the longitudinal wrinkles has a more complex relationship to applied tension than previously estimated. The clamped edges are shown to suppress longitudinal wrinkling over a distance set by the applied tension and the ribbon width, but otherwise have no apparent effect on measured wavelength. Further, by analyzing the stress profile, we find that longitudinal wrinkling does not completely alleviate compression, but caps the magnitude of the compression. Nonetheless, the width over which wrinkles form is observed to be wider than the near-threshold analysis predictions-- the width is more consistent with the predictions of far-from-threshold analysis. However, the end-to-end contraction of the ribbon as a function of twist is found to more closely follow the corresponding near-threshold prediction as tension in the ribbon is increased, in contrast to the expectations of far-from-threshold analysis. These results point to the need for further theoretical analysis of this rich thin elastic system, guided by our physically robust and intuitive simulation model.
\end{abstract}

\maketitle

\section{Introduction}
Twisting a thin ribbon under tension generates compression, somewhat counterintuitively, and consequently longitudinal wrinkles in the center of the ribbon. Larger twist angles lead to the even more whimsical creased helicoid phase, sometimes referred to as ``ribbon crystals’’. These nonintuitive deformation modes hint at deep physics, and despite twisting being the basis of ancient textile technologies---from twisting fibers into yarns to coiling ropes into piles---the mechanics of twisted morphologies remain hazy~\cite{hearle1969structural,Pan2002,bohr_2011}. These mysteries have unfurled slowly over decades, beginning when Green first observed buckling and wrinkling patterns in ribbons and analyzed the development of compression with twist~\cite{green_elastic_1937}. It was not until nearly fifty years later that the existence of a buckling transition in twisted plates was confirmed numerically~\cite{crispino_stability_1986}, and it took another twenty years to verify numerically the longitudinal wrinkling pattern Green described~\cite{coman_asymptotic_2008}. Shortly after, the creased helicoid phase was modeled geometrically in the isometric limit~\cite{korte_triangular_2011, bohr_ribbon_2013}.
Next, experimental probes provided a full map of the twisted ribbon phase space, revealing a sprawling zoo of helicoids, wrinkles, and loops which are dependent on twist angle and applied tension~\cite{chopin_helicoids_2013}; analytical attempts to characterize these post-buckling phenomena quickly followed~\cite{chopin_roadmap_2014,Kohn2018}. Surprisingly, creased helicoids still formed in ribbons with tension, developing from the wrinkles themselves~\cite{chopin_disclinations_2016} and displaying some amount of stretching in the ridges (unseating the isometric assumptions of previous studies)~\cite{pham_dinh_cylindrical_2016}. Ribbons of finite-thickness are analytically slippery, requiring approximations whose regions of validity not fully clear~\cite{chopin_roadmap_2014,chopin_extreme_2019}. To fully resolve the limitations of these approximations requires data such as internal energy and stress distributions, quantities which are currently difficult to access by experiment or theory.

In general, thin sheets are excellent candidates for computational modeling. Plenty of great work has been done to simulate thin sheet deformations using finite element methods (FEM)~\cite{bridson_robust_2002,narain_folding_2013,schreck_nonsmooth_2016,gottesman_localized_2018,charrondiere_numerical_2020}, and there are several mass-spring-models that successfully map regular discrete lattices to the bulk properties of a sheet~\cite{seung_defects_1988,ostoja-starzewski_lattice_2002,plummer_buckling_2020}. Less work, however, has been dedicated to mapping a microscopically random mesh's parameters to the bulk properties of the simulated sheet. Some models use a random mesh to approximate a constant bulk Young's modulus~\cite{van_Gelder_1998,lloyd_identification_2007}, and others focus on making discretized bending realistic~\cite{grinspun_discrete_2003,wardetzky_discrete_2007,tamstorf_discrete_2013}. But in some cases these models are inconsistent with analytical descriptions of regular meshes; what's more, a combined stretching and bending model has to our knowledge, not been thoroughly tested. A discrete mesh model that compares directly to physical materials is infinitely useful, opening the door to generating large data sets useful for data-driven discoveries.

In this paper, we develop a simple, computationally cheap, mechanical mass-spring-model (MSM) to study twisted thin ribbons. We could equivalently use an FEM approach to study thin twisted ribbons. However, we are compelled by the MSM because it is an intuitive extension of coupled oscillators; allows local, ``microscopic'' control of the mesh topology; and \revb{is not tethered to the limitations of using a partial differential equation for the sheet, such as describing the post-buckled shapes, or singularities in the PDE in areas with stress-focusing}. It is a simple model with historical precedence that accurately replicates the interesting observables of the thin elastic sheet, \revb{using only Hookean springs with small stresses that depend linearly on deformation}. A lattice model like this, with nearest and next-to-nearest neighbor interactions, also preserves the possibility of learning about or from other statistical mechanics lattice models.

\begin{figure*}
    \centering
    \includegraphics[width=17cm]{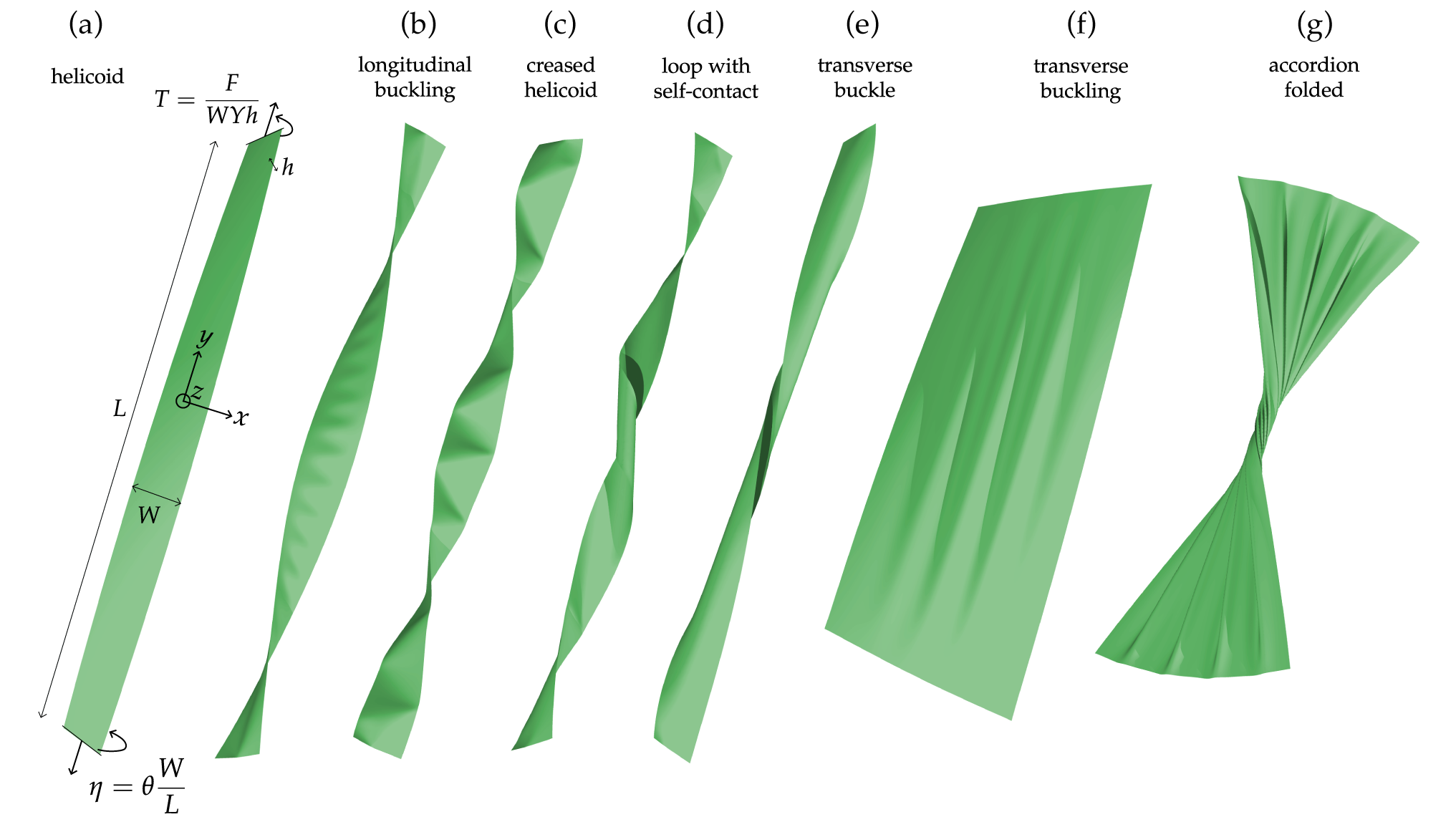}
    \caption{Deformation modes of twisted thin sheets, replicated by our simulation whose details are summarized in Fig.~\ref{fig:model}. Ribbon (a) is labeled with its dimensions and quantities relevant to the twisting-under-tension procedure. $T$ is the scaled longitudinal tension applied, which is total applied force $F$ normalized by the width, Young's modulus, and thickness of the ribbon, and $\eta$ is a scaled twist angle: the end-to-end angle $\theta$ normalized by the ribbon's aspect ratio. These quantities are defined in Table \ref{tab:variables}. (a) The helicoid phase initially present before any buckling transitions occur. (b) Longitudinal buckling occurs at an angle $\eta_\text{lon}$ and has a fixed wavelength $\lambda_\text{lon}$ set by the tension $T$ and thickness $h$. (c) Creased helicoids develop from the longitudinally buckled ribbons as the buckle ridges ``turn'' to form triangular facets. (d) At tensions below the crossover tension $T^*$, ribbons will snap-through to form a loop at an angle $\eta_\text{tran}$. If twisted far enough, ribbons will develop self-contact at an angle $\eta_\text{sc}$. (e) Transverse buckling occurs at tensions greater than $T^*$, transitioning at angle $\eta_\text{tran}$. (f) The wavelength of transverse buckling $\lambda_\text{tran}$ is set by the aspect ratio of the sheet; sheets with smaller aspect ratios can display several wavelengths of transverse buckling. (g) At high twists, low aspect ratio sheets display ``accordion folding'' and approach the yarning transition~\cite{chopin_tensional_2022}. Snapshots (a)--(e) fall within the phase diagram presented in Fig.~\ref{fig:phase_diagram}.}
    \label{fig:phases}
\end{figure*}

Our randomly-seeded MSM maps reliably to a physical Young's modulus and bending rigidity, allowing us to match experimental conditions and generate the various modes of deformations observed as a function of ribbon aspect ratio, applied tension and twist (see Fig.~\ref{fig:phases}). We are able to carefully analyze the onset and growth of wrinkles in the longitudinally buckled mode using measurements which are also available experimentally, such as surface curvature. Additionally, the simulations provide access to the invisible internal dynamics of the ribbon, including finely resolved spatial and temporal data for the strain and stress. These new insights are useful for characterizing the mechanics of the buckling transition, and probing the relevance of existing near- and far-from-threshold approximations. Guided by recent rounds of physical observations and analytical inquiries, we show that numerics and simulations once again hold the key to the next stage of discovery about twisted thin ribbons.

\section{Mass Spring Model}

\begin{figure}[h!]
    \centering
    \includegraphics[width=8.4cm]{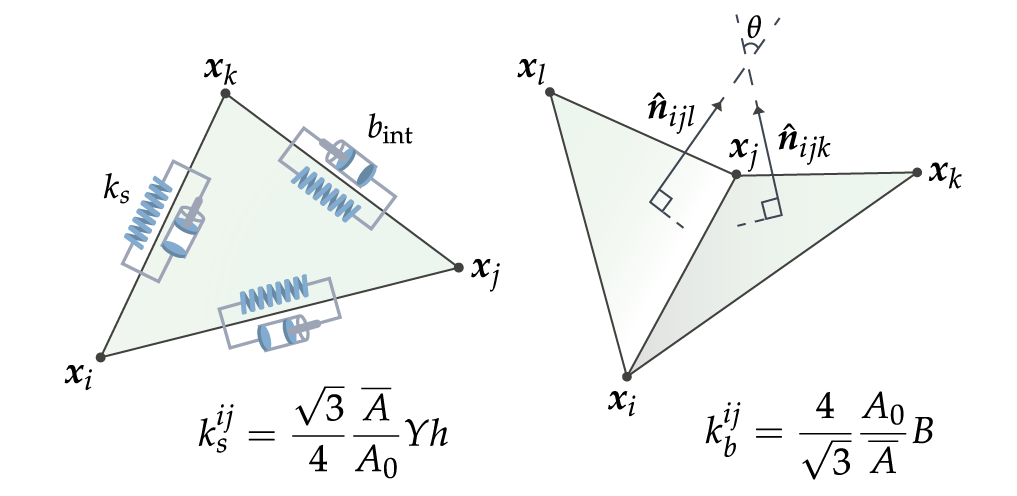}
    \caption{Nodes are connected by in-plane springs with stiffness $k_s$ and dashpot damping $b_\text{int}$. The $k_s$ of each in-plane spring is set by the local geometry of the mesh and the target Young's modulus for the sheet, $Y$. Out-of-plane stiffness $k_b$ is similarly dictated by the local geometry and the target bending rigidity, $B$. Both sets of springs ensure quadratic energy penalization for stretching and misalignment of the vectors normal to the triangular mesh facets, Eqs.~\eqref{eq:rand_stretch_eng} and \eqref{eq:rand_bend_eng}.}
    \label{fig:model}
\end{figure}

We model a thin ribbon by defining a mesh with a set of nodes, arranged either in a regular triangular lattice (where each interior node has six equidistant nearest neighbors) or a random triangular lattice in which node coordination number and distance to nearest neighbors vary. Nodes are connected to their neighbors via in-plane springs and dashpots, and bending is controlled by pseudo-springs (a quadratic penalization of bending) across each edge between adjacent triangles, as shown in Fig.~\ref{fig:model}. \revb{A full description of this discrete model's relationship to continuum elasticity is provided in Appendix~\ref{appendix:model}; in this section we provide a brief history of the model and a summary of the modifications we employ.}

The regular triangular lattice is well-studied and oft-utilized, with the attractive feature of having an analytical mapping from the discrete stretching and bending spring constants ($k_s$ and $k_b$) to a continuous two-dimensional Young's modulus and bending rigidity for the sheet. The springs $\bm{r}_{ij}$ connecting pairs of lattice sites, $\bm{x}_{i}$ and $\bm{x}_j$ dictate in-plane deformations. Hinges separating two triangles $\triangle_{ijk}$ and $\triangle_{ikl}$, with normal vectors $\hat{\bm{n}}_{ijk}$ and $\hat{\bm{n}}_{ikl}$, control out-of-plane motion. Seung and Nelson~\cite{seung_defects_1988} showed that for the triangular lattice with unit length springs, and potentials of the form
\begin{equation}
\label{eq:stretch_eng}
    E_s\left(\bm{r}_{ij}\right) = \frac{1}{2} k_s \left(1 - \left|\bm{x}_i - \bm{x}_j\right|\right)^2,
\end{equation}
\begin{equation}
\label{eq:bend_eng}
    E_b \left(\hat{\bm{n}}_{ijk}, \hat{\bm{n}}_{ikl}\right) = \frac{1}{2} k_b \left|\hat{\bm{n}}_{ijk} - \hat{\bm{n}}_{ikl} \right|^2,
\end{equation}
the equivalent continuous two-dimensional (2D) Young's modulus and bending rigidity for the sheet are
\begin{equation}
\label{eq:moduli}
Y_{2D} = \frac{2}{\sqrt{3}} k_s, \enskip B = \frac{\sqrt{3}}{2}k_b,
\end{equation}
respectively. If the rest configuration of the lattice deviates from hexagonal packing of equilateral triangles (such as at the boundaries of a rectangular sheet or a collection of randomly placed lattice points), the above relationships are no longer \revb{correct beyond the leading order discrete approximation of the continuum}. To extend this mass-spring-model to irregular lattice configurations, it is useful to modify the pre-factor of each energy term.

Each in-plane spring represents an area of continuous material that resists stretching or compression. Thus springs adjacent to larger triangular facets should have stiffer spring constants to reflect the greater amount of material they represent. We modify the stretching energy term according to Van Gelder~\cite{van_Gelder_1998} (with typos in the original model corrected by Lloyd et al.~\cite{lloyd_identification_2007}):
\begin{equation}
    \begin{split}
    \label{eq:rand_stretch_eng}
        E_s(\bm{r}_{ij}) &= \frac{1}{2} \left(\frac{1}{2} \frac{\overline{A}}{A_0} k_s\right)\left(s_{ij} - \left|\bm{x}_i - \bm{x}_j\right|\right)^2, \\ &= \frac{1}{2} \left(\frac{\sqrt{3}}{4} \frac{\overline{A}}{A_0}  Y_{2D}\right) \left(s_{ij} - \left|\bm{x}_i - \bm{x}_j\right|\right)^2,
    \end{split}
\end{equation}
where $Y_{2D}$ is the target Young's modulus for the sheet, $s_{ij}$ the rest length of a given spring, $\overline{A}$ the sum of the facet areas adjacent to edge $\bm{r}_{ij}$, and $A_0$ the area of an equilateral triangle with side length $s_{ij}$. When the entire lattice is composed of equilateral triangles, this expression reduces to the model in Eq.~\eqref{eq:stretch_eng}.

With similar physical motivations as given above\revb{---namely that an area of continuous material distributed over a longer length scale should be floppier, or easier to bend---}we modify the bending energy term to be
\begin{equation}
    \begin{split}
    \label{eq:rand_bend_eng}
        E_b\left(\hat{\textbf{n}}_{ijk}, \hat{\textbf{n}}_{ikl}\right) &= \frac{1}{2} \left(2 \frac{A_0}{\overline{A}} k_b \right) \left| \hat{\textbf{n}}_{ijk} - \hat{\textbf{n}}_{ikl} \right|^2 \\ &= \frac{1}{2} \left(\frac{4}{\sqrt{3}} \frac{A_0}{\overline{A}} B \right) \left| \hat{\textbf{n}}_{ijk} - \hat{\textbf{n}}_{ikl}\right|^2.
    \end{split}
\end{equation}
The area dependence of the coefficient is inspired by Grinspun et al.~\cite{grinspun_discrete_2003} and adapted such that it reduces to the Seung and Nelson bending energy (Eq.~\eqref{eq:bend_eng}) when all triangles are equilateral. \revb{Although this quantity is presented as a ratio of areas, it implicitly accounts for the shapes of the adjacent triangles through the quantity $\overline{A}$. Grinspun et al.~\cite{grinspun_discrete_2003} provide a full explanation of this shape consideration.}

Although our mesh is 2D, the bending rigidity imparts an effective thickness to the sheet. The bending rigidity $B$ is related to the Young's modulus $Y$ by~\cite{landaulifshitz}
\begin{equation}
    \label{eq:bending_rigidity}
    B = \frac{Y h^3}{12(1-\nu^2)},
\end{equation}
where $h$ is the thickness of the material and $\nu$ is Poisson's ratio. For triangular lattices $\nu$ cannot be independently tuned, so $\nu = 1/3$ always~\cite{seung_defects_1988,lloyd_identification_2007}, and $Y = Y_{2D}/h$. Thus we can rearrange Eq.~\eqref{eq:bending_rigidity} and use the relationships in Eq.~\eqref{eq:moduli} to define an effective thickness:
\begin{equation}
    \label{eq:eff_thickness}
    h_{\text{eff}} = \sqrt{\frac{8 k_b}{k_s}}\, .
\end{equation}

This length scale is used to set the sheet's self-avoidance: the sheet is not allowed to come within $h_{\text{eff}}$ of itself. Self-avoidance is enforced by introducing a repulsive force between contact sites within an interaction range $h_{\text{eff}}$ of each other~\cite{andrejevic_simulation_2022}. In general, the contact sites are spaced more closely than the mesh nodes. For example, the thinnest sheet we consider here ($h =\SI{127}{\micro\meter}$) has a mesh spacing about 8 times coarser than $h_\text{eff}$. Thus we implement ``level 3'' refinement (iterative bisection of triangle edges three times, placing additional contact sites at the midpoints) such that the spacing of contact sites is on the order of the sheet's thickness. Forces between the refined contact sites are distributed to the nodes of the mesh nearest to the sites, weighted by proximity to the site. The site refinement applies only during contact detection and allows the mesh to remain coarse in all other calculations. From now on, mentions of the simulated sheet's thickness $h$ are in reference to this effective thickness.

\begin{table}[h]
    \centering
    \renewcommand{\arraystretch}{1.5}
    \begin{tabular}{>{\raggedright}p{0.43\linewidth}|>{\centering}m{0.15\linewidth}|>{\centering\arraybackslash}m{0.32\linewidth}}
         Property\,  & \,Symbol\, & \,Formula \\
         \hline
         Length & $L$ & $-L/2 < y < L/2$ \\
         Width & $W$ & $-W/2 < x < W/2$ \\
         Thickness & $h$ & \\
         Young's modulus & $Y$ & $Y_{2D}/h$\\
         Poisson's ratio & $\nu$ & $1/3$  \\
         Bending rigidity & $B$ & $\frac{Y h^3}{12\left(1-\nu^2\right)}$ \\
         End-to-end twist angle & $\theta$ & \\
         Scaled twist angle & $\eta$ & $\theta \frac{W}{L}$ \\
         Scaled applied tension & $T$ & $ \frac{F}{Y h W}$ \\
         Confinement parameter & $\alpha$ & $\eta^2/T$
    \end{tabular}
    \caption{Definitions of variables and physical parameters.}
    \label{tab:variables}
\end{table}

The equations of motion for a node $i$ in the sheet are
\begin{align}
    \begin{split}
        \dot{\bm{x}}_i &= \bm{v}_i, \\
        m \bm{a}_i &= \bm{F}_i,
    \end{split}
    \label{eq:eom}
\end{align}
where $\bm{x}_i$ is the three-dimensional position of a node, $\bm{v}_i$ its velocity, \revb{$\dot{\bm{v}}_i = \bm{a}_i$} its acceleration, and $\bm{F}_i$ the sum of the forces applied to node $i$, which is given by
\begin{align}
  \bm{F}_i &= -\sum_{j \in N_i} \nabla_{\bm{x}_i} \left(E_s\left(\bm{r}_{ij}\right)\right) - \sum_{\substack{(ijk)\in T_i \\ (ikl) \in T_i}} \nabla_{\bm{x}_i}\left(E_b\left(\hat{\bm{n}}_{ijk}, \hat{\bm{n}}_{ikl}\right)\right) \nonumber \\
           &\phantom{=}+ \sum_{j \in N_i} \bm{F}_d^\text{int}\left(\bm{x}_i,\bm{x}_j, \bm{v}_i, \bm{v}_j \right) + \bm{F}_d^\text{iso}\left(\bm{v}_i\right) \nonumber \\
           &\phantom{=}+ \sum_{j} \bm{F}_c \left(\bm{x}_i, \bm{x}_j\right) + \bm{F}_\text{ext}.
\end{align}
Here $N_i$ are the neighbors of node $i$, and $T_i$ are the adjacent triangles. The damping term $\bm{F}_d^\text{int}$ is due to the dashpots pictured in Fig.~\ref{fig:model}, and $\bm{F}_d^\text{iso}$ is an additional isotropic, global damping. $\bm{F}_c$ is the contact force, which is repulsive and turns on when two nodes come within $h$ distance of one another~\cite{andrejevic_simulation_2022}. Finally, any external applied forces are included in $\bm{F}_\text{ext}$.

High-frequency elastic waves dissipate quickly in materials that are of interest to us (e.g.\@ paper, Mylar, aluminum foil). Instead, wrinkling is characterized by slower deformations on the scale of the system size, with occasional hysteretic, snap-through events. We therefore assume the sheet is mostly in a quasistatic regime where the acceleration of a node is close to zero: $\bm{a}_i \approx 0$. The quasistatic equations of motion constitute a differential-algebraic system which is well-suited to an implicit numerical integration scheme in time~\cite{andrejevic_simulation_2022}. When the quasistatic approximation is violated, such as near a snap-through event accompanied by large changes in node velocities, the full dynamic equations of motion given in Eq.~\eqref{eq:eom} are integrated explicitly instead. Further details on the numerical integration schemes and switching criteria are provided in Appendix~\ref{appendix:numerics}, and a thorough explanation is given in Ref.~\cite{andrejevic_simulation_2022}.

We note that plasticity can be added to this model by allowing the rest angle (length) of a hinge (spring) to change if the hinge (spring) is deformed past a specified yield threshold. The plastic damage can then accumulate according to a purely plastic, strain hardening, or even strain weakening model~\cite{andrejevic_simulation_2022}.
In this work, however, we will discuss only purely elastic sheets which do not fatigue or accumulate damage.

\section{End-to-End Twisting}

\subsection{Ribbon Setup and Boundary Conditions}

Our simulated ribbons correspond to length $L = \SI{45.72}{\centi\metre}$ and width $W= \SI{2.54}{\centi\metre}$. Three different thicknesses are used throughout this paper: \SI{127}{\micro\metre}, \SI{254}{\micro\metre}, and \SI{508}{\micro\metre}. A single, randomly-seeded ribbon mesh was used across all simulations, with an average node spacing of $d=\SI{1}{\milli\meter}$. The random lattice is generated by seeding the ribbon with nodes using the Voro++ library~\cite{rycroft_voro_2009,lu22}, then Lloyd's algorithm~\cite{lloyd1982} is iteratively applied 100 times to make the mesh more regular. The mesh configuration at this point is taken as its rest configuration, and all the springs and facets have differing lengths and areas. The spring and hinge stiffnesses are set according to the model in Eqs.~\eqref{eq:rand_stretch_eng} and \eqref{eq:rand_bend_eng}, and tuned such that the Young's modulus of the sheet is approximately $Y = 3.40 \, \text{GPa}$ and the bending rigidity is approximately $B =\SI{.653}{\milli\pascal\cdot\meter^3}$, $B =\SI{5.22}{\milli\pascal\cdot\meter^3}$, or $B = \SI{41.8}{\milli\pascal\cdot\meter^3}$, respectively, for the three thicknesses. Properties of the thinnest ribbon are given in physical units in Table \ref{tab:props}.

\begin{table}[h!]
    \centering
    \renewcommand{\arraystretch}{1.5}
    \begin{tabular}{>{\raggedright}p{0.43\linewidth}|>{\centering}m{0.15\linewidth}|>{\centering\arraybackslash}m{0.32\linewidth}}
        Property\, & \,Symbol\, & \,Measurement\, \\
        \hline
        Length & $L$ & $\SI{45.7}{\centi\meter}$ \\
        Width & $W$ & $\SI{2.54}{\centi\meter}$ \\
        Thickness & $h$ & $\SI{127}{\micro\meter}$ \\
        Young's modulus & $Y$ & $\SI{3.40}{\giga\pascal}$ \\
        Bending rigidity & $B$ & $\SI{0.653}{\milli\pascal \cdot \meter^3}$ \\
        Poisson ratio & $\nu$ & 0.333 \\
    \end{tabular}
    \caption{Properties of the primary test ribbon, in physical units.}
    \label{tab:props}
\end{table}

For both the regular and randomly seeded meshes we performed a series of three modulus convergence tests: stretch, shear, and bend. The average node spacing had the range $d \in [\SI{0.5}{\milli \meter}, \SI{2.5}{\milli\meter}]$. In Appendix~\ref{appendix:mod_tests} we see that the model in Eqs.~\eqref{eq:rand_stretch_eng},~\eqref{eq:rand_bend_eng} converges to the expected values of Young's modulus ($Y$), shear modulus ($G$), and bending rigidity ($B$). While there is some amount of error in these moduli, it is within the range of variation expected, for example, from physical material samples.

As in most of the twisted ribbon experimental setups~\cite{green_elastic_1937,chopin_helicoids_2013,bohr_ribbon_2013,pham_dinh_cylindrical_2016}, the short edges of our ribbon are clamped such that the nodes are fixed in a rigid line. The clamped edges then rotate with relative angular velocity $\dot{\theta} = 0.15$ rad/s (up to time $t = \SI{120}{\second} = \SI{2}{\minute}$) to produce a net end-to-end twist angle $\theta$ across the ribbon. This rate is sufficiently slow for the loading to be considered quasi-static. This angle is then normalized by the aspect ratio to give the scaled twist angle $\eta = \theta \frac{W}{L}$. Additionally, a weight $F$ (in Newtons) is applied longitudinally to the clamped edges. This force is then scaled by the ribbon's dimensions and Young's modulus such that $T = \frac{F}{Y h W} \in [1.80 \times 10^{-4}, 4.14 \times 10^{-3}]$. In each simulation the ribbon was twisted at least until it reached a transverse buckling point; most of the the final angle twists were between $\eta \in [0.5, 0.9]$. The computation time per simulation ranged from 1--4 days running on 10--14 threads, with the longer runs mostly depending on the significant contact refinement needed for very thin sheets that encounter self-contact.

\subsection{Deformation Modes}

\begin{figure}
    \centering
    \includegraphics[width=8.6cm]{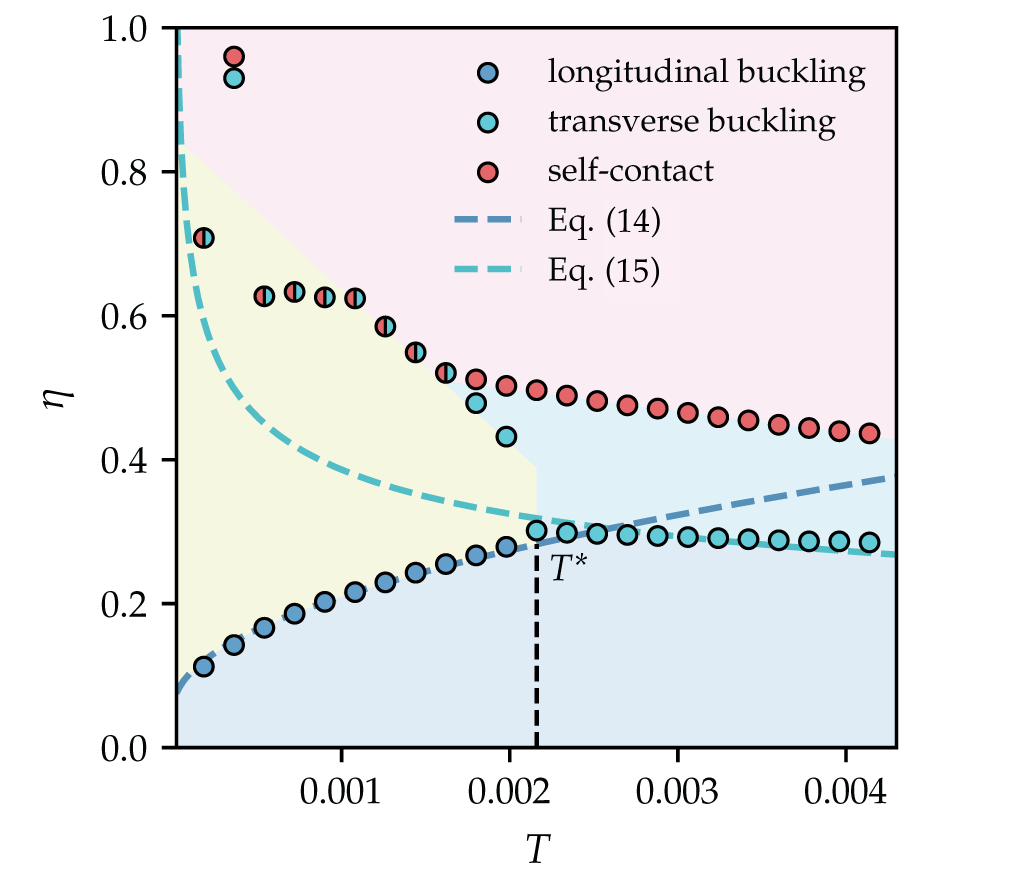}
  \caption{A slice of the thin ribbon deformation phase space at fixed thickness $h = \SI{127}{\micro\meter}$. Plotted are longitudinal buckling transitions (blue circles), transverse buckling transitions (green circles), and points of self-contact (red circles) as a function of scaled ribbon tension $T$ and scaled twist angle $\eta$. At low tensions the transverse buckling transition often leads immediately to self-contact, whereas at higher tensions the transverse instability occurs before self-contact develops. The blue dashed line corresponds to the theoretical scaling of the longitudinal buckling angle $\eta_\text{lon}$ (Eq.~\eqref{eq:lon_buckling}) with a single fitting parameter for the coefficient of the finite thickness correction term. The green dashed line is the theoretical scaling of the transverse buckling angle $\eta_\text{tran}$ (Eq.~\eqref{eq:tran_buckling}), again with a single fitting parameter for the coefficient. $T^*$, the tension at which the primary instability switches from longitudinal to transverse buckling, is indicated by the black dashed line.}
    \label{fig:phase_diagram}
\end{figure}

At least seven distinct deformation modes have been experimentally observed~\cite{chopin_helicoids_2013,chopin_roadmap_2014,chopin_tensional_2022}. Inspired by the variety of these modes, and the careful tuning necessary to find them all, we set out to test our computational model's ability to reproduce all the physically observed deformations using our simulated ribbons. By adjusting only the total load applied to the ribbon along the longitudinal axis ($T$) and the twist angle ($\eta$), we indeed find all the reported experimentally observed morphologies, as displayed in Fig.~\ref{fig:phases}.

We initially see the pure helicoid phase, Fig.~\ref{fig:phases}a, which transitions to the longitudinally buckled ribbon, Fig.~\ref{fig:phases}b, at an angle $\eta_\text{lon}$. The creased helicoid, Fig.~\ref{fig:phases}c, develops from this longitudinally buckled phase. Creased helicoids can then undergo a looping transition, Fig.~\ref{fig:phases}d, at $\eta_\text{tran}$ and will eventually develop self-contact at $\eta_\text{sc}$. At greater tensions we observe transitions from the helicoid to the transverse-buckled sheet, Fig.~\ref{fig:phases}e. Transverse-buckled ribbons can also reach self-contact at $\eta_\text{sc}$. Sheets with smaller aspect ratios develop multiple wavelengths of transverse buckling, Fig.~\ref{fig:phases}f, and can enter the accordion folded/yarning regime, Fig.~\ref{fig:phases}g.

Overall we see excellent qualitative and quantitative agreement of our simulations with experiments, from visual inspection of the various instabilities to a thorough recreation of a slice of the twisted ribbon phase space, in Fig.~\ref{fig:phase_diagram}. With these practical tests and the modulus convergence tests (Young's, shear, and bend) in Appendix \ref{appendix:mod_tests}, we have full confidence in our simulations' ability to replicate physical sheets and the nuances of their deformations. We find that some subtle ambiguities that were raised in experiments-- such as the nature of longitudinal wrinkling's onset and spectrum of wrinkling frequencies, as well as dependence of the buckling wavelength $\lambda_\text{lon}$ on $h$ and $W$-- are revealed more clearly through the precise measurements possible via our simulations.

\subsubsection{Phase Space}
The ribbon's deformation modes are arranged in a three-dimensional (3D) phase space with axes of sheet thickness $h$, normalized twist angle $\eta$, and normalized longitudinal load $T$ (note that other dimensional and physical properties such as aspect ratio and Young's modulus are absorbed in the normalized quantities, explained further in Table \ref{tab:variables}). By fixing the thickness, one can take a 2D slice of this phase space with the remaining variables being the applied load and the twist angle.

We have fixed the thickness at $h = \SI{127}{\micro\meter}$ in order to recreate a slice of the ribbon deformation phase space that was probed experimentally by Chopin et al.\@~\cite{chopin_helicoids_2013}. Throughout the slow twist, we extract the angles $\eta$ at which the primary instability, any secondary instabilities, and moments of self-contact occur. The primary instability could be either longitudinal or transverse buckling ($\eta_{\text{lon}}$ or $\eta_{\text{tran}}$); in the former case the ribbon will undergo a secondary transverse buckling instability at a higher twist angle ($\eta_{\text{tran}}$). If twisted far enough, ribbons will reach self-contact ($\eta_{\text{sc}}$). All of these transitions are plotted in Fig.~\ref{fig:phase_diagram}, and the longitudinal instability scales with a dependence on $T$ and an offset related to the ribbon's thickness, presented in Eq.~\eqref{eq:lon_buckling}. The transverse instability also has a dependence on $T$, and additional dependences on the dimensions of the ribbon, shown in Eq.~\eqref{eq:tran_buckling}. The crossover tension $T^*$ indicates the primary instability switching from longitudinal buckling to transverse buckling. (Modes shown in Fig.~\ref{fig:phases}a-e fall within this slice of the phase space, whereas the varied aspect ratio and thickness of modes Fig.~\ref{fig:phases}f,g place these examples in different slices of the 3D phase space.)

\begin{figure*}
    \centering
    \includegraphics[width=17cm]{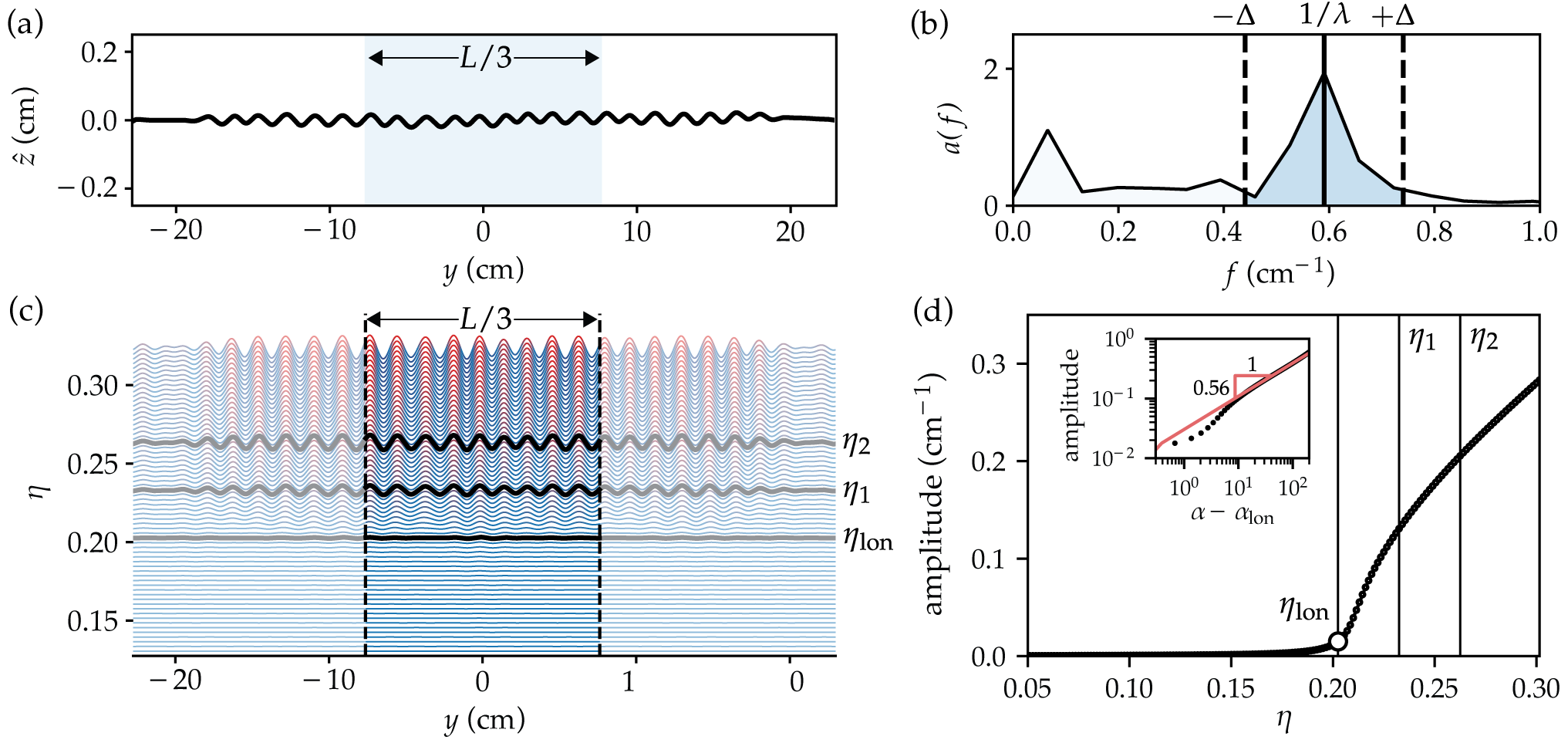}
    \caption{Detection of longitudinal wrinkles. (a) The out-of-plane displacement $\hat{z}$ along the midline of a sample ribbon \SI{127}{\micro\metre} thick with scaled tension $T=9.00\times{10}^{-4}$, which undergoes longitudinal buckling. Note that due to the clamped boundary conditions the wrinkles are suppressed a distance $L_\text{sup} \sim \mathcal{O}\left(W\right)$.
    (b) The amplitude spectrum $a(f)$ of the middle third of the wrinkle profile in (a), computed as the absolute value of the discrete Fourier transform of $\hat{z}$. The frequency of the highest peak is taken to be the reciprocal of the wrinkle wavelength $\lambda$. To aid in identifying the precise onset of wrinkling, the amplitude spectrum is truncated to retain only frequencies in the range $\left(1/\lambda\right) \pm \Delta$, and the wrinkle profile is reconstructed from the inverse Fourier transform of the truncated spectrum. (c) A waterfall plot of reconstructed wrinkle profiles as a function of scaled twist angle $\eta$ along the $y$-axis, for which frequencies $\left(1/\lambda\right)\pm\SI{0.15}{\per\centi\metre}$ have been retained, with $\lambda\approx\SI{1.69}{\centi\metre}$. (d) The wrinkle amplitude $(\langle H^2(r=0)\rangle_y)^{1/2}$ as a function of $\eta$, where the average $\left<\cdot\right>_y$ is computed over the middle third of the wrinkle profiles in (c). The three identified scaled angles $\eta_\text{lon}$, $\eta_1$, and $\eta_2$ correspond to selected wrinkle profiles in (c) shown in black. The scaled angle $\eta_\text{lon}={0.203}$ marks the detected onset of longitudinal wrinkling, identified as the ``knee point'' ~\cite{satopaa2011finding} of the amplitude curve. Inset: Amplitude as a function of confinement parameter offset by the wrinkle onset, $\alpha-\alpha_{\text{lon}}$, with scaling $\beta=0.56$ (Eq.~\eqref{eq:amp_scaling}).}
    \label{fig:wrinkling_analysis}
\end{figure*}

The onset of longitudinal wrinkles is determined by the out-of-plane displacement along the midline of the ribbon, as illustrated in Fig.~\ref{fig:wrinkling_analysis}a. After the onset of wrinkling, we observe that the frequency content of the wrinkle profile is peaked at a characteristic frequency $f=1/\lambda$, with $\lambda$ being the dominant wrinkle wavelength (Fig.~\ref{fig:wrinkling_analysis}b). To enhance the detection of wrinkle onset, frequencies outside a specified range near the characteristic frequency are removed from the wrinkle profile, as shown in Fig.~\ref{fig:wrinkling_analysis}b--c. The resulting wrinkle amplitude (see Fig.~\ref{fig:wrinkling_analysis} caption for definition) exhibits a sharp increase at a scaled twist angle $\eta_{\text{lon}}$; this ``knee point'' of the amplitude curve\reva{---approximately the point of maximum curvature in the data~\cite{satopaa2011finding}---}is the onset of longitudinal buckling (Fig.~\ref{fig:wrinkling_analysis}d).

\begin{figure*}
    \centering
    \includegraphics[width=17cm]{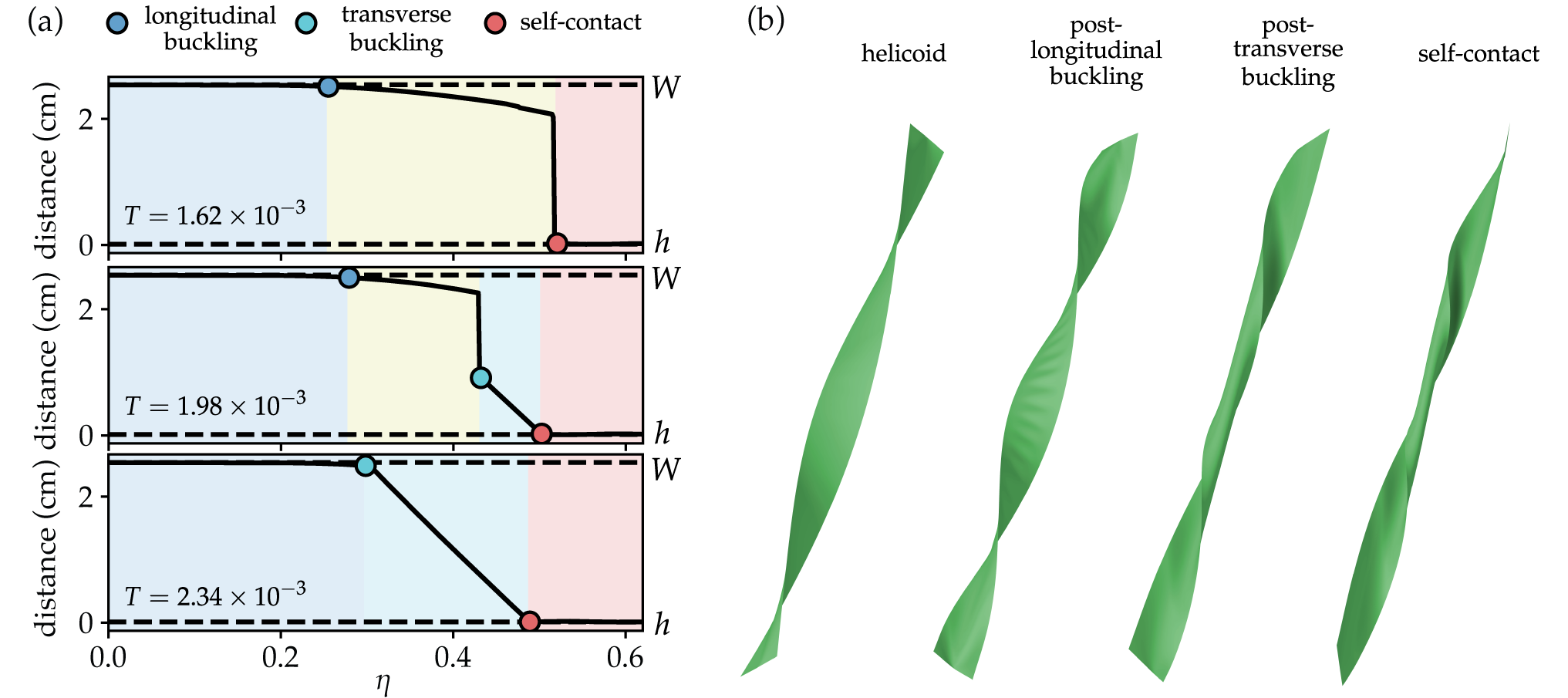}
    \caption{Detection of transverse buckling and self-contact. (a) The shortest distance between any two points on opposing long edges of a ribbon as a function of scaled twist angle, shown for ribbons at three different scaled tensions. At $\eta=0$, the edge distance is equal to the ribbon width $W$, while after self-contact, the edge distance equals the effective thickness $h$. Solid markers indicate the onset of longitudinal buckling, transverse buckling, and self-contact, mapping out a vertical trajectory through the phase space plot of Fig.~\ref{fig:phase_diagram} along constant $T$. (b) Snapshots of the ribbon in (a) (middle) at $T=1.98\times{10}^{-3}$ in each of the four deformation regimes.}
    \label{fig:buckling_contact}
\end{figure*}

Transverse buckling and self-contact are detected by measuring the shortest distance between any two points on opposing long edges of a ribbon. At $\eta = 0 $ this distance is equal to the ribbon width $W$. However, as $\eta$ increases, we continue to track this distance, as shown in Fig.~\ref{fig:buckling_contact}a. The onset of transverse buckling is marked by a pronounced change in the slope of the edge distance plot. A second abrupt change in slope marks the point of self-contact of the ribbon, after which the shortest distance between opposing long edges is approximately constant near the ribbon's effective thickness.

While the transverse buckling and self-contact transitions observed for large twist and tension are rich and fascinating phenomena~\cite{mockensturm_elastic_2000,chopin_helicoids_2013,chopin_roadmap_2014,kudrolli_tension-dependent_2018,chopin_tensional_2022}, for the remainder of this work we will focus on the small $T$, small $\eta$ region. Buckling transitions in this corner of the phase diagram have been well-observed and categorized experimentally~\cite{chopin_helicoids_2013}, but near-threshold and far-from-threshold analytical efforts to describe these instabilities are challenged by a lack of closed form solutions to describe the post-buckling ribbon shapes~\cite{crispino_stability_1986,coman_asymptotic_2008,chopin_roadmap_2014}. The approximations available are expected to be valid only within a very narrow window $\eta\gtrsim \eta_{\text{lon}}$, even in the so called far-from-threshold analysis, \revb{where the analysis is based on a variant of the FvK equations, which is weakly non-linear in the small amplitude ``out-of-ribbon'' displacements~\cite{chopin_roadmap_2014}.} Our simulations enable a critical examination of the nature of the transitions.

\subsubsection{Wavelength Scaling}
\begin{figure}
    \centering
    \includegraphics[width=8.6cm]{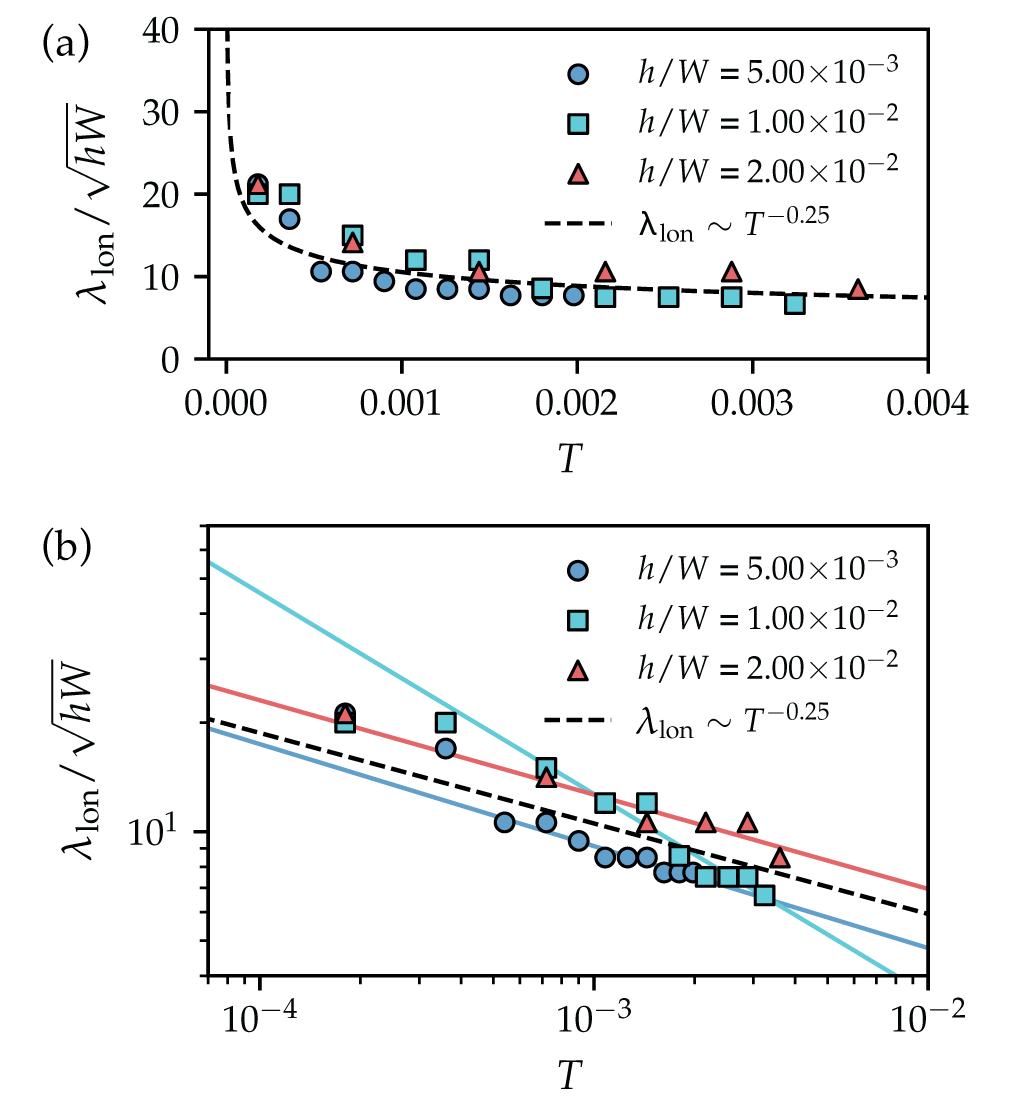}
    \caption{Variation of scaled longitudinal wavelength $\lambda_{\text{lon}}/\sqrt{hW}$ with scaled tension $T$. (a) The wavelength $\lambda_{\text{lon}}$ of the wrinkles was estimated in experiments to scale as $T^{-1/4}$, and good agreement is obtained by fitting a curve through the data at all but the two smallest tensions ($T<4\times{10}^{-4}$), as shown by the dashed line~\cite{chopin_helicoids_2013}. Prior experimental results also reveal deviations from the theoretical scaling at low tensions~\cite{chopin_helicoids_2013}. (b) Individual fits of scaled wavelength for each of the three thickness to width ratios $h/W$ (solid lines) shown on a log--log scale, with the theoretical scaling reproduced from (a) (dashed line). A closer examination of the data from (a) reveals that $\lambda_\text{lon}$ has a lingering dependence on the thickness $h$, and could have a more complex relationship to $T$ than previously estimated. }
    \label{fig:wavelength}
\end{figure}

The primary instability in the small $\eta$, small $T$ corner of the phase space is longitudinal buckling (Fig.~\ref{fig:phases}b). The spine of the ribbon is plotted as a function of longitudinal position in Fig.~\ref{fig:wrinkling_analysis}a; when the ribbon buckles, a pattern of wavelength $\lambda_\text{lon}$ appears along the center of the ribbon.

The scaling of $\lambda_\text{lon}$ with the initial tension $T$ was derived analytically by Coman and Bassom~\cite{coman_asymptotic_2008} with singular perturbation methods of the second-order boundary value problem, and by Chopin and Kudrolli~\cite{chopin_helicoids_2013} using energy scaling arguments. More sophisticated scaling arguments noting the relation of wavelength to wrinkling width were also derived later by Chopin, et al ~\cite{chopin_roadmap_2014}. These methods of obtaining a scaling relation for the wavelength $\lambda_\text{lon}$, gave rise to the same scaling
\begin{align}
\label{eq:wavelength}
    \frac{\lambda_{\text{lon}}}{\sqrt{hW}} \propto T^{-1/4} \,,
\end{align}
where all made use of the near-threshold (NT) approximation, which assumes a small amplitude for the wrinkles. The NT analysis makes no claims to describe the wrinkling pattern as $\eta$ and the amplitude of the wrinkles increases, but is a fine reference point for analyzing the onset of the longitudinal buckling phase.

Figure \ref{fig:wavelength}a demonstrates that if the wavelength $\lambda_\text{lon}$ is normalized by the width and thickness, the data points tend to collapse on the curve $T^{-1/4}$. These simulated results agree with previous experimental measurements of wavelength~\cite{chopin_helicoids_2013}. However, we find that ribbons held at low tensions deviate from the observed $T^{-1/4}$ dependence, indicating other parameters might be subdominant at moderate tensions but important in the small $T$ limit. A closer look at the data in log--log space, Fig.~\ref{fig:wavelength}b, reveals the points' scatter systematically depends on the ribbon thickness. Thus, our simulations show that  $\lambda_\text{lon}$ has a more complicated dependence on $h$, and potentially $T$, than previously estimated.

\subsubsection{Wrinkle Suppression}

\begin{figure}
    \centering
    \includegraphics[width=8.6cm]{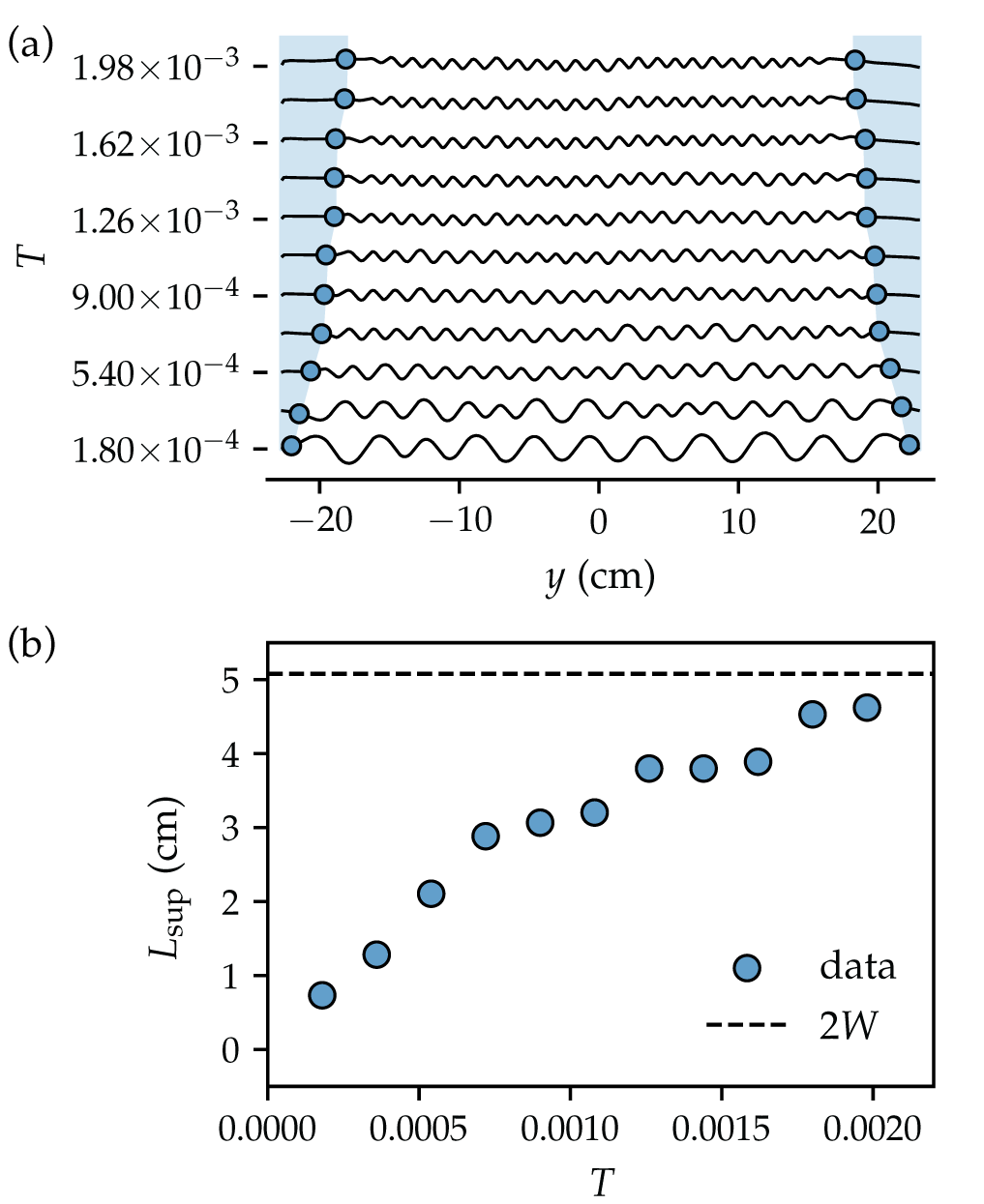}
    \caption{Wrinkle suppression in a ribbon with $h = \SI{127}{\micro\meter}$ due to clamped boundaries. (a) Profiles of all longitudinally buckled samples, highlighting the wrinkle suppression zone near the ribbon edges in blue. The suppression distance measured from either clamp is $L_\text{sup}$. (b) Wrinkle suppression length, $L_\text{sup}$ as a function of applied tension $T$. The suppression near a clamped edge approaches the value $2W = \SI{5.08}{\centi\meter}$ as tension increases toward $T^*$. $L_\text{sup}$ likely depends on $h$ as well as $W$.}
    \label{fig:l_supp}
\end{figure}

The clamped boundary conditions suppress the wrinkles by a distance of $L_\text{sup}~\sim~\mathcal{O}(W)$ from each edge. $L_\text{sup}$ is measured from our simulation data as the distance from the clamp at which the wrinkle amplitude exceeds $5\%$ of its value at an angle $\eta > \eta_{\text{lon}}$ when wrinkles are well-developed (equivalent to $\eta_2$ indicated in Fig.~\ref{fig:wrinkling_analysis}c--d). The wrinkle amplitude is defined as in Fig.~\ref{fig:wrinkling_analysis}d, and is reflected over the $y=0$ line and averaged to determine a single average value for $L_\text{sup}$ at each tension. Fig.~\ref{fig:l_supp}a displays the profile of each longitudinally buckled sample. The regions shaded blue are defined as the suppression zone near the clamped boundary. As shown in Fig.~\ref{fig:l_supp}b, the suppression distance is very small at low tensions and increases to nearly $2W$ at greater tensions, close to $T^*$ (the upper bound for longitudinal wrinkling).
Despite this suppression near the edges, the wavelength $\lambda_\text{lon}$ of the wrinkles is not otherwise affected by the clamped boundaries. We base this on the fact that the buckled wavelength is consistent everywhere outside of the suppression zone, both in the middle of the ribbon and near the clamps.

\subsection{Mechanical Responses}
\label{subsection:mechanics}

A wide variety of information is accessible to us through our simulations, such as built-in x-ray vision which allows us to track the distribution of strain and stress in the sheet throughout the simulated experiment. These insights allow us to perform powerful analyses of the sheets' deformations, and examine details of the twisted ribbon problem which have been previously intractable through experimental means. In particular, we investigate the detailed evolution of the ribbon's stress distribution, wrinkling, and the contraction as it deforms and undergoes the longitudinal buckling instability at low tensions.

\subsubsection{Primary Instabilities}

\begin{figure}
    \centering
    \includegraphics[width=8.6cm]{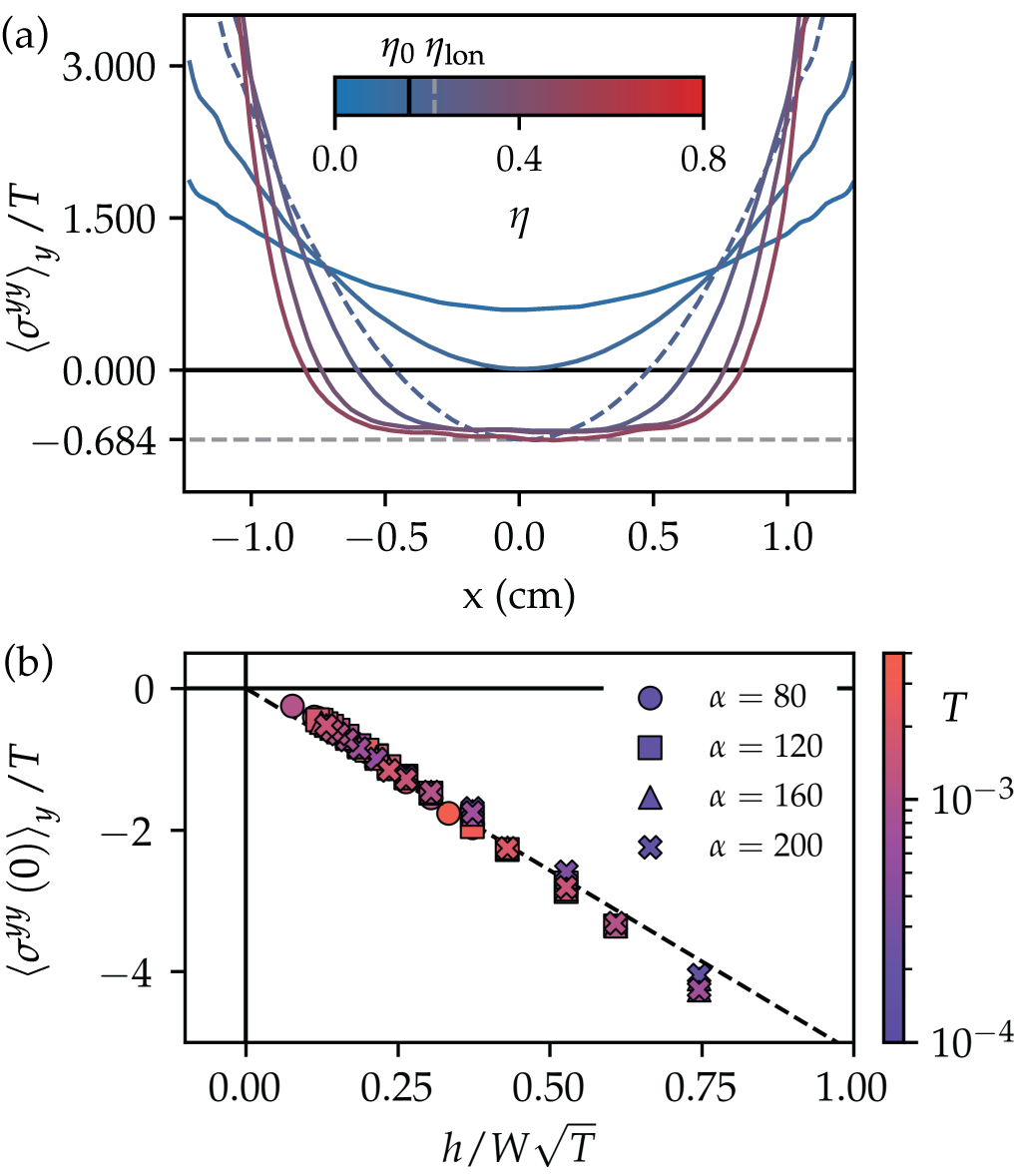}
    \caption{(a) Stress profile of a ribbon with $T = 1.08 \times 10^{-3}$, $h = \SI{127}{\micro \meter}$. The color bar (and color of each stress slice) indicates the progression of the ribbon's twist, with blue corresponding to $\eta = 0$ and red $\eta = 0.8$, which is the fully twisted state for this sample. The solid black line in the color bar marks $\eta_0$, when stress becomes compressive, and the gray dashed line marks $\eta_\text{lon}$, the onset of longitudinal buckling. As the ribbon twists, its longitudinal stress profile is initially parabolic, crossing over to compressive stress in the center of the ribbon at $\eta_0$ ($= 0.161$). After the longitudinal buckling transition at $\eta_\text{lon}$ ($= 0.216$, dashed blue-purple line) the stress profile bottoms out (minimum stress marked by the horizontal gray, dashed line) and widens. The buckled ribbon thus continues to support compressive stress, but functionally caps the compressive load. (b) At varying thicknesses ($h = \SI{63.5}{\micro \meter}, \SI{127}{\micro \meter}, \SI{254}{\micro \meter}, \SI{508}{\micro \meter}$) and degrees of confinement ($\alpha = 80, 120, 160, 200$), we see that the residual stress supported by the longitudinally buckled ribbon is linear in $h/(W\sqrt{T})$, with a slope of $-5.14\pm0.05$ given by the dashed line. Regardless of $\alpha$, as the thickness tends to zero we expect the stress in the ribbon to also vanish.}
    \label{fig:stress}
\end{figure}

The transition from pure helicoid to longitudinal or transverse buckling can be predicted analytically from the dimensionless stress profile of a twisted ribbon~\cite{chopin_roadmap_2014}:
\begin{align}
        \sigma^{yy}(x) &= T + \frac{\eta^2}{2} \left( \left(\frac{x}{W}\right)^2 -\frac{1}{12}\right) \label{eq:lon_stress}\\
        \sigma^{xx}(x) &= \frac{\eta^2}{2} \left(\left(\frac{x}{W}\right)^2 - \frac{1}{4}\right) \left[T + \frac{\eta^2}{4} \left(\left(\frac{x}{W}\right)^2 + \frac{1}{12}\right) \right].
    \label{eq:tran_stress}
\end{align}
The ribbon supports a smooth twist in the form of a helicoid for small $\eta$. However, as $\eta$ increases, depending on the applied load $T$, the longitudinal  stress $\sigma^{yy}$ \reva{of the centerline ($x=0$)} becomes compressive (negative) at an angle $\eta_0$:
\begin{align}
    \eta_0= \sqrt{24T} \, .
\end{align}
In an infinitely thin ribbon, any amount of compressive stress would cause the ribbon to buckle, but a ribbon of finite thickness can withstand an amount of compression proportional to its thickness, generating a corrective factor~\cite{chopin_helicoids_2013},
\begin{align}
\label{eq:lon_buckling}
    \eta_{\text{lon}} = \sqrt{24 T} + C_\text{lon} \frac{h}{W} \,.
\end{align}
Here $C_\text{lon} = 11.00$ is a fitting parameter extracted from the longitudinal buckling transitions plotted in Fig.~\ref{fig:phase_diagram}, which is close to the experimentally measured $C_\text{lon} = 9.3$ for ribbons with similar dimensions~\cite{chopin_helicoids_2013}.

\begin{figure*}
    \centering
    \includegraphics[width=17.2cm]{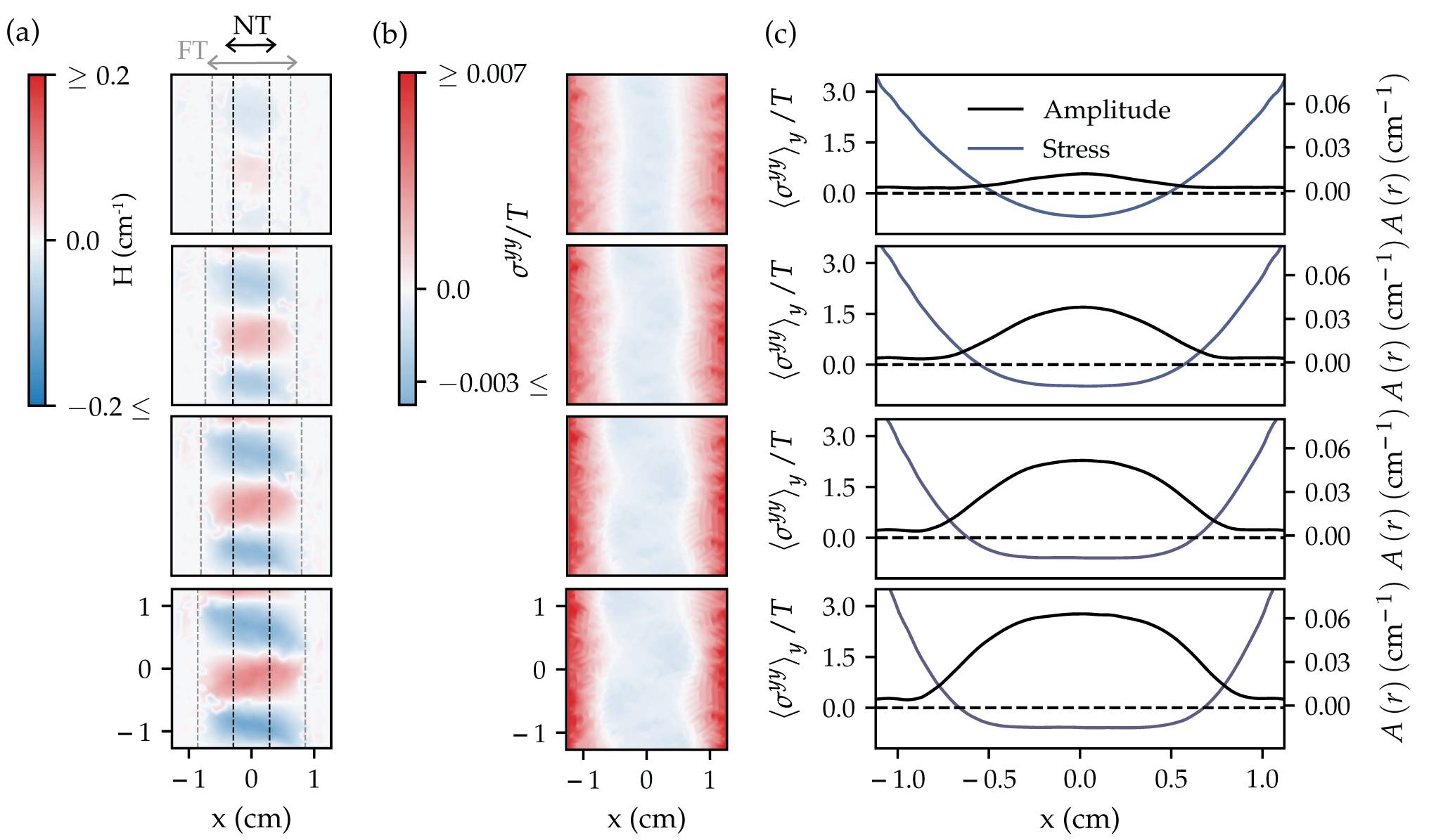}
    \caption{Development of the wrinkles that appear in a ribbon with a tension $T = 1.08 \times 10^{-3}$, $h = \SI{127}{\micro\meter}$. Snapshots were taken at regular intervals, starting at the onset of buckling. \reva{The first frame is at $\eta = \eta_{\text{lon}}$, and the frames below show the progression of the wrinkles as the ribbon goes through the longitudinally buckled phase and begins to cross into the creased helicoid phase.} (a) Contour maps of the mean curvature in the ribbon. Red (blue) corresponds to above(below)-the-plane curvature. Black (gray) dashed lines mark the near threshold (far-from-threshold) predictions for the width of the wrinkles. At the onset of wrinkling their width is consistent with the NT prediction, but they soon expand to meet the FT prediction. In the final frame we also see the wrinkle ridges begin to ``turn'' and start to form the triangular facets characteristic of the creased helicoid phase. \reva{(b) Contour maps of the longitudinal stress component. Red (blue) corresponds to tensile (compressive) in-plane stress. As the twist angle increases, the magnitude of the compressive stress remains approximately the same, though the width of the compressive area broadens. The transition from longitudinal buckling to the angled creased helicoid is also evident in these stress maps.} (c) The stress profile and resultant amplitude ($A = (\langle H^2(r)\rangle_y)^{1/2}$) at various values of $\eta$.
    }
    \label{fig:amplitude}
\end{figure*}

The expected transverse buckling angle is estimated by stress balancing scaling arguments~\cite{chopin_roadmap_2014}. In ribbons of finite length (i.e.\@ not extremely long), the transverse buckling angle is
\begin{align}
\label{eq:tran_buckling}
    \eta_\text{tran} = C_\text{tran} \sqrt{\frac{h}{L}} T^{-1/4} \, ,
\end{align}
where $C_\text{tran}$ is a dimensionless constant.
We compare this form with the transverse buckling transitions we observed in Fig.~\ref{fig:phase_diagram}, and find that $C_\text{tran} = 4.12$ describes
transitions that occur at $T>T^*$.

\subsubsection{Stress Distribution}

Moving forward we will discuss only the longitudinal stress, as the transverse stress components are suppressed within the small $\eta$, small $T$ region we study here. A quantity that  will be useful in forthcoming calculations is the ``confinement parameter'' $\alpha = \eta^2/T$, which allows us to compare the progression of twist across samples with different applied tensions. It can also be understood as the ratio of geometric to tensile strain~\cite{chopin_roadmap_2014}.

For $\eta>\eta_0$, the compressive stress occurs in a symmetric region about the longitudinal center line (i.e.\@ $x/W<r_{\text{wr}}$ with $2r_{\text{wr}}$ being the characteristic width of the compression zone), as seen by solving Eq.~\eqref{eq:lon_stress}. The profile will remain parabolic until the twist reaches the critical buckling angle, $\eta_\text{lon}$. This parabolic profile is that of a pure helicoid; the shape of the stress profile at $\eta>\eta_\text{lon}$, however, will reveal the nature of the longitudinal buckling transition and which, if either, of the existing analytical approximations (near-threshold or far-from-threshold) can describe the post-buckling ribbon~\cite{chopin_roadmap_2014}. \revb{The parameters of our primary test ribbon ($T \in \left[1.8 \times 10^{-4}, 2 \times 10^{-3}\right)$ and $h = \SI{127}{\micro\meter}$) place it in a regime which could be considered near-threshold (NT), or might be in an ambiguous region between the NT and far-from-threshold (FT) zones.} If the stress continues to be parabolic for some time post-buckling, \revb{then we would see that the longitudinally buckled ribbon is well described by the NT analysis, and within the region of NT validity. On the other hand, the FT procedure assumes the compressive stress in the wrinkled zone is zero at a first order expansion about a compression-free state. Ribbons with finite thickness are expected to support some compressive stress,} in which case we would expect $\sigma^{yy}(x<r_{\text{wr}}) = 0$, \revb{or at least to decrease substantially,} when $\eta>\eta_\text{lon}$.

By analyzing the local stress tensors at each facet of our mesh, we find that for a given applied load $T$ ($=1.08 \times 10^{-3}$), the stress profile of our simulated ribbon (Fig.~\ref{fig:stress}a) becomes compressive at precisely the $\eta_0 = \sqrt{24 T}$ ($= 0.161$) that was predicted by Eq.~\eqref{eq:lon_stress}. In Fig.~\ref{fig:stress}a the longitudinal buckling transition occurs at $\eta_{\text{lon}} = 0.216$, after which the stress profile begins to flatten out. The shape of this profile seems to approach the theoretical profile for an infinitely thin ribbon in the FT approximation~\cite{chopin_roadmap_2014}. Importantly, however, we find that the buckled ribbon continues to support compressive stress, and the width of the compressive zone widens as the twist progresses. This was predicted by Qiu~\cite{qiu_morphology_2017}, but \revb{analytically computing what the residual stress should be for the twisted ribbon remains challenging and has not yet been done in the literature.}

\revb{In Fig.~\ref{fig:stress}b we plot the residual stress at various values of $\alpha$ and across ribbons of varying thickness. Regardless of the degree of confinement, the ribbon's residual stress is linear in $h/(W\sqrt{T})$. Longitudinal buckling functionally caps the magnitude of stress the ribbon supports-- we can see this clearly by the collapse of data points onto the same fitted line in Fig.~\ref{fig:stress}b. At a fixed thickness and tension the residual stress is the same, even when the value of $\alpha$ increases. As the thickness of the ribbon tends to zero, we expect the stress to vanish, since an infinitely thin ribbon cannot support compression. This intuition is also supported by our data.}

\reva{Residual stresses are seen in other systems~\cite{vella_indentation_2015,hohlfeld_sheet_2015,paulsen_curvature-induced_2016}; a saturation of residual stress has even been predicted for a circular sheet confined to a curved substrate~\cite{davidovitch_geometrically_2019}. However, given that the twisted ribbon is not externally confined or adhered to a substrate, the relation between the ribbon's residual stress and these other examples is not immediately obvious.} The residual stress arises for the same reason $\eta_{\text{lon}}$ exceeds $\eta_0$: the non-zero thickness of the ribbon~\cite{chopin_helicoids_2013}. Like a beam, a ribbon of finite thickness can support some amount of compression. Conversely, some amount of compression seems necessary to maintain the buckling pattern, as the lack of stress would allow the ribbon to relax back into a flat state. Despite lacking a precise explanation for the measured phenomenon, these results are a crucial insight to the nature of the longitudinal wrinkling transition: namely, this transition is completely reversible since no hysteresis occurs, and the longitudinally buckled and creased helicoid phases support compression in the center of the ribbon.

\subsubsection{Wrinkle Amplitude and Confinement}

Transverse slices of the longitudinal wrinkles can be extracted from a map of the mean curvature:
\begin{equation}
\label{eq:amp}
    A(r) = \sqrt{\left<H^2(r)\right>_y}\,.
\end{equation}
Note that in this definition of amplitude, the mean curvature $H(r)$ is averaged along the longitudinal ($y$) axis in the central third of the ribbon. In Fig.~\ref{fig:amplitude}a are plots of $H(r)$, with red (blue) zones indicating above-the-plane (below-the-plane) curvature. As the twist progresses the wrinkles store more curvature and also begin to ``turn'' into the triangular facets indicative of the creased helicoid phase. Figure~\ref{fig:amplitude}b plots Eq.~\eqref{eq:amp} versus the transverse position, as well as the corresponding stress profile (like the slices in Fig.~\ref{fig:stress}a) measured at the same $\eta$ (reminiscent of a position wavefunction and its potential well). The edges of the ribbon are in tension, so they remain flat, whereas the center of the ribbon contains the longitudinal wrinkles which form to almost fully relieve the compression.

\begin{figure}
    \centering
    \includegraphics[width=8.6cm]{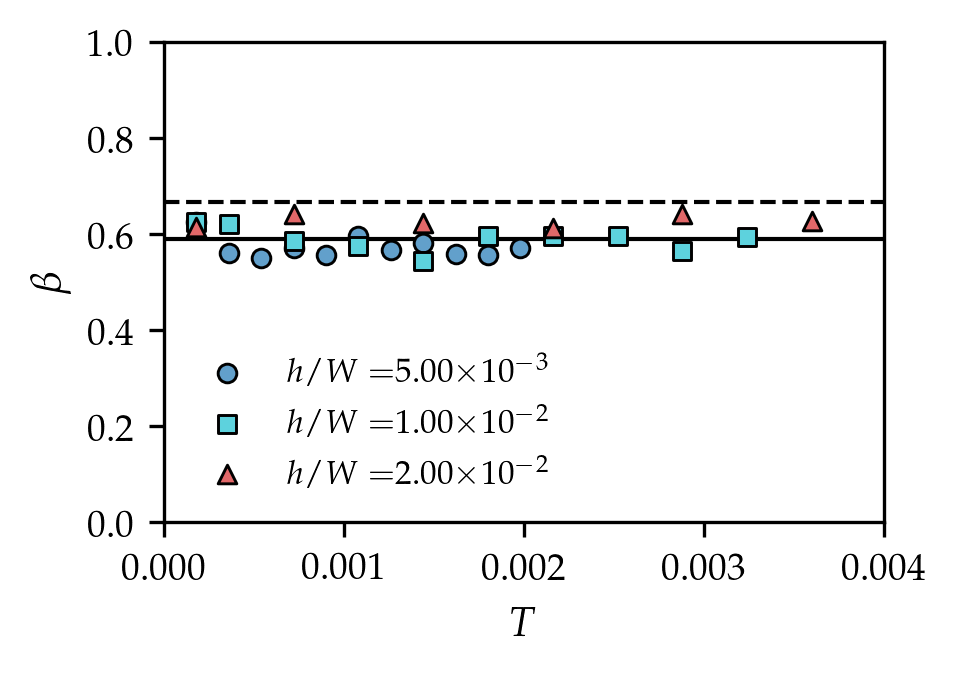}
    \caption{The maximum amplitude of longitudinal wrinkles scales like $A(r = 0) \propto \left(\alpha - \alpha_\text{lon}\right)^\beta$. Values of $\beta$ are extracted from the amplitude curve in Fig.~\ref{fig:wrinkling_analysis}d and are plotted here for ribbons of varying thickness and applied tension. We find $\beta = 0.59\pm0.05$ with the average value given by the solid black line. The horizontal dashed black line is the value Chopin and Kudrolli~\cite{chopin_disclinations_2016} experimentally estimated to be $\beta = 2/3$. }
    \label{fig:beta}
\end{figure}

\begin{figure*}
    \centering
    \includegraphics[width=17.2cm]{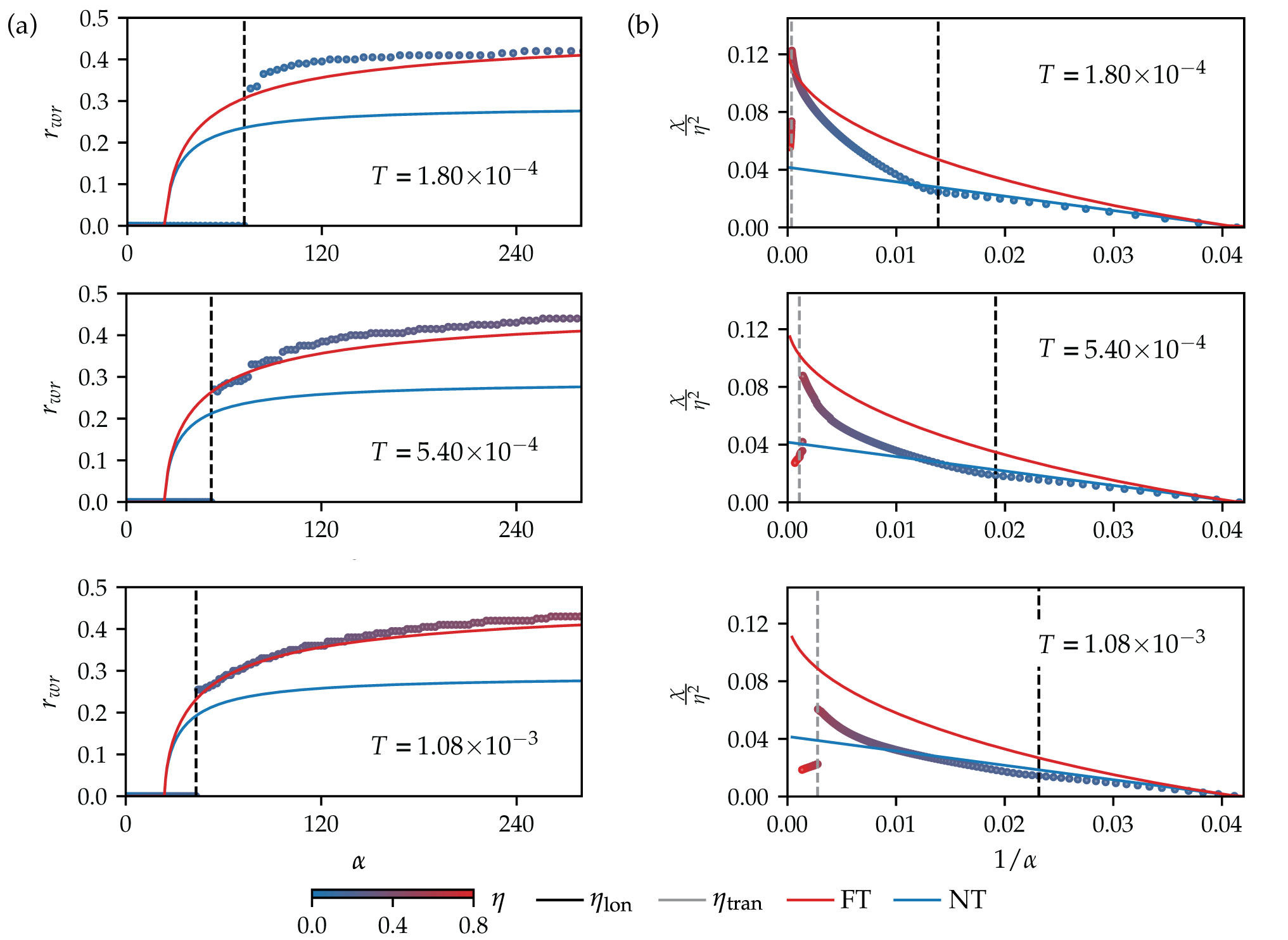}
    \caption{Tests of the NT and FT model predictions at $h = \SI{127}{\micro\meter}$, various ribbon tensions. (a) At the onset of wrinkling ($\eta_{\text{lon}}$, dashed black line) the wrinkled zone emerges at a non-zero width. It widens as the twist progresses, eventually asymptoting to $r_{\text{wr}} = 0.5$ as $\alpha \to \infty$. This behavior is similar across various applied tensions, though at moderate tensions (e.g.\@ $T = 1.08 \times 10^{-3}$) $r_{\text{wr}}$ adheres more closely to the theoretical FT prediction than it does at lower tensions. (b) Another way to test the NT and FT predictions is by measuring the contraction $\chi$ of the ribbon. Here $\chi$ is plotted against the inverse confinement parameter (such that small values of $1/\alpha$ correspond to the greatest angles $\eta$). Across varying tensions the ribbon tends to contract as a pure helicoid, following the NT prediction, until the onset of wrinkling ($\eta_{\text{lon}}$, dashed black line). Post wrinkling, the contraction proceeds rapidly, at times even faster than the predicted FT slope. This difference in scaling could be because of the transition to the creased helicoid phase, which is predicted to have a different contraction scaling than the FT model~\cite{korte_triangular_2011,pham_dinh_cylindrical_2016,chopin_extreme_2019}. The gray dashed line corresponds to the transverse buckling point, after which the ribbon springs back to a lesser contraction.}
    \label{fig:ntvft}
\end{figure*}

As the twist progresses, the maximum amplitude $A(r=0)$ gradually increases. We identify $\eta_{\text{lon}}$, the onset of longitudinal wrinkles, as the angle corresponding to the \reva{the ``knee points'', the approximate points of maximum curvature~\cite{satopaa2011finding},} in the $\eta$ vs amplitude curve, identified in Fig. \ref{fig:wrinkling_analysis}d. At angles larger than $\eta_{\text{lon}}$, the magnitude of the amplitude continues to grow according to
\begin{align}
    A \sim (\alpha - \alpha_{\text{lon}})^\beta.\label{eq:amp_scaling}
\end{align}
(Recall that $\alpha = \eta^2 / T$.) The exponent $\beta$ is extracted from the \reva{sizeable linear portion of the} slope of the amplitude curve in log--log space, shown in the inset of Fig.~\ref{fig:wrinkling_analysis}d. Fig.~\ref{fig:beta} shows the extracted $\beta$ from the fits as a function of $T$ for the three different ribbon thickness probed in our simulation. We find that $\beta =0.59\pm 0.05$ is approximately constant, albeit with some scatter which may suggest some dependence of $\beta$ on the thickness of the sheet: the amplitude of the wrinkles increases more rapidly (larger value of $\beta$) for thicker ribbons. This is most likely because thicker ribbons have longer buckling wavelengths $\lambda_\text{lon}$, with fewer total ridges fitting along the length of the ribbon. The $\beta$ found in our simulations is similar to that reported by Chopin and Kudrolli~\cite{chopin_disclinations_2016} based on experimental measurements which had more limited precision. While analytical calculations are currently lacking, we hope that our results will lead to further work in this direction.

We measure the width of the wrinkled zone $r_{\text{wr}}$ by identifying ``knee'' of the amplitude profile (Fig.~\ref{fig:amplitude}b), and plot this $r_{\text{wr}}$ as a function of $\eta$ in Fig.~\ref{fig:ntvft}a. Chopin et al.~\cite{chopin_roadmap_2014} predicted that for a ribbon of finite thickness, $r_{\text{wr}} = 0$ until the critical $\eta_{\text{lon}}$ after which $r_{\text{wr}}$ might jump directly to the NT curve (\cite{chopin_roadmap_2014}, Eq.~(3)),
\begin{align}
    \left(1 - 12 r_\text{wr}^2\right) = \frac{24}{\alpha},
\end{align}
and eventually progress to following the FT curve (\cite{chopin_roadmap_2014} Eq.~(50)),
\begin{align}
    \left(1 - 2 r_\text{wr}\right)^2\left(1+4r_\text{wr}\right) = \frac{24}{\alpha}
\end{align}
Our simulation agrees with this in spirit, though we observe that after longitudinal buckling (denoted by the vertical dashed black line) the ribbon jumps directly to the FT prediction (plotted in solid red), similar to experimental observations with ribbons with similar thickness~\cite{chopin_disclinations_2016}. We also find that $r_{\text{wr}}$ grows a bit faster than the FT prediction. Moderate tensions, such as $T = 1.08 \times 10^{-3}$, seem to more closely follow the FT prediction than smaller tensions. Using a thinner ribbon and a very fine temporal resolution might show wrinkles developing at widths near the NT curve (plotted in solid blue) before widening to the FT curve.

\subsubsection{Length Contraction}

Another metric that distinguishes the NT from FT approximations is $\chi = 1 - L_\text{ee}/L_0$, the contraction of the ribbon where $L_\text{ee}$ is the end-to-end ribbon length at a given $\eta$, and $L_0$ is the ribbon length at $\eta_0 = \sqrt{24T}$~\cite{pham_dinh_cylindrical_2016}. Fig.~\ref{fig:ntvft}b plot $\chi/\eta^2$ as a function of inverse confinement ($1/\alpha$). Chopin et al. further summarize that as a ribbon is twisted, a pure helicoid will contract as (\cite{chopin_roadmap_2014}, Eq.~(20)),
\begin{align}
    \frac{\chi_{\text{NT}}}{\eta^2} &= \frac{1}{24} - \frac{1}{\alpha}\, ,
\end{align}
(plotted in solid blue) whereas a ribbon in the FT approximation would contract according to (\cite{chopin_roadmap_2014}, Eq.~(52)),
\begin{align}
    \frac{\chi_\text{FT}}{\eta^2} &= \frac{r_\text{wr}^2}{2}\,,
\end{align}
(plotted in solid red). For three simulations with various applied tension, we find that initially the ribbon does contract as a helicoid. Then immediately after the onset of wrinkles it begins to contract more quickly, similar to the FT predictions. At times, however, the contraction grows even quicker than the slope of the FT curve, particularly at low tensions. This difference in scaling could be because of the ribbon's transition into the creased helicoid phase, which contracts more as an isometric packing of the triangular facets than as an elastic wrinkling problem (i.e.\@ the FT prediction)~\cite{korte_triangular_2011,pham_dinh_cylindrical_2016,chopin_extreme_2019}.

It has been suspected that the NT approximations well describe the onset of longitudinal buckling, and that the FT approach is useful at larger twist angles; the transition between these regimes is admittedly fuzzy~\cite{chopin_roadmap_2014,chopin_extreme_2019}. NT analysis does a great job at predicting $\lambda_\text{lon}$~\cite{coman_asymptotic_2008}, and we show throughout this section that FT analysis is able to capture much of the post-buckling behavior, such as the general shape of the stress profile and width of the wrinkled zone.
We find that despite their individual successes, neither the NT or FT analyses can entirely capture the onset or development of the longitudinal wrinkling phase. This deficiency is shown most clearly in the contraction of the ribbon.

\section{Conclusion}

We have used simple, computationally cheap, mass-spring-model simulations to recreate the rich morphology
and phase behavior of twisted ribbons, which were previously realized experimentally and analyzed theoretically.
The subtleties of studying twisted ribbon morphology provided the perfect test playground for our mechanical model of thin sheets. We chose a mass-spring-model as a simple, nostalgic extension of the ubiquitous coupled oscillators physical system. The fine spatial control of the MSM provides intuition for the mesoscopic physics (much coarser than atomic but still discrete approximation of the continuum) which drives morphological transitions in elastic sheets. This particular MSM has a long history, with analytical mappings for continuous bulk properties when the underlying lattice is regular~\cite{seung_defects_1988}. Attempts have been made in the past to generalize the model to be useful for a random lattice~\cite{van_Gelder_1998,lloyd_identification_2007,grinspun_discrete_2003,wardetzky_discrete_2007,tamstorf_discrete_2013}, and we propose a hybrid of these past attempts which we have shown translates the discrete mesh parameters to the continuous bulk properties, regardless of underlying mesh topology.

Through precise transition detection and wrinkling analysis we have thoroughly probed the small twist, small applied tension regime. We carefully examined the scaling of the longitudinal buckling wavelength, and observed a region of wrinkle suppression near the clamped edges. As twist increases, the wrinkle amplitude grows with a robust scaling constant $\beta = 0.59\pm0.05$. Furthermore, studying the stress profile of the ribbon  revealed that longitudinal buckling caps the amount of compression the ribbon supports with its finite thickness. Measurements of the wrinkled zone width and the ribbon's net contraction additionally reveal that the near- and far-from-threshold approximations are able to capture some, but not all, of the ribbons' behavior.

These simulations are useful for probing regimes of the twisted ribbon phase space which are difficult to study experimentally. Additional deformation modes are suspected to theoretically lurk in the low tension and very thin ``corners'' of the phase space~\cite{chopin_roadmap_2014}, approaching the tensionless and isometric limits. \revb{These two regimes are straightforward to study using simulations, since the applied tension and thickness of the ribbon can be set arbitrarily small in our framework. A slightly different computational approach is necessary for the exactly isometric case: a differential-algebraic equation (DAE) solver is necessary to impose additional algebraic constraints on the system. Our method already uses a DAE solver, which could be readily adapted to study isometric sheets with our methods.} One can also imagine treating the transverse buckling and looping transitions with the same attention we have devoted to longitudinal buckling, including the charming yarning transition~\cite{kudrolli_tension-dependent_2018}. The torque response of ribbons during the twist procedure could also be extracted from simulations and compared to experiments~\cite{liu_novel_2022}. The development of e-cones and d-cones (highly localized deformation and stress focusing) in the creased helicoid phase is another topic which could be further illuminated through simulations~\cite{farmer_geometry_2005,witten_stress_2007,chopin_disclinations_2016}. Deeper understanding of these myriad fascinating instabilities remains to be unlocked by computational studies.

Further applications of this simulation framework include deformations of curved films or shells, wrinkling on substrates, flat-folding and origami, and crumpling through various geometries. Meshes can be generated with any boundary shape or cutouts, and the resultant sheet can be pre-stressed or plastically deformed (e.g.\@ dimpling). Our mesh model is well-suited for studying mechanical responses of a sheet (pre- and post-deformation), and internal stress and energy measurements allow many modes of data collection and analysis.

\section{Acknowledgements}

We thank Julien Chopin and Benjamin Davidovitch for their expert thoughts and discussions. This research was partially supported by NSF through the Harvard University Materials Research Science and Engineering Center DMR-2011754. ML was supported by the Ford Foundation Predoctoral Fellowship and both ML and JA were supported by the National Science Foundation Graduate Research Fellowship Program under grant no. DGE-1745303. AK was supported by National Science Foundation grant DMR-2005090. CHR was partially supported by the Applied Mathematics Program of the U.S.\@ DOE Office of Science Advanced
Scientific Computing Research under contract number DE-AC02-05CH11231.

\bibliography{ref}

\appendix

\section{Discretized model of a continuous sheet}\label{appendix:model}
\subsection{Stretching}
\revb{In modeling a continuous sheet, we choose a discretization scheme inspired by the Seung and Nelson (SN) model~\cite{seung_defects_1988} for a mesh of equilateral triangles. We extend the SN model to apply to meshes with triangles of varying shape by modifying the prefactors of the energy terms. Beginning with the in-plane stretching, the energy density of a continuous elastic sheet can be written as
\begin{equation}
    u_s = \frac{1}{2} \epsilon \cdot C \cdot \epsilon
\end{equation}
where $\epsilon$ is the in-plane strain tensor and $C$ is the stiffness tensor. A 2D, isometric material has a stiffness tensor of the form
\begin{equation}
    \begin{bmatrix}
    C_{1111} & C_{1122} & C_{1112}\\
    C_{2211} & C_{2222} & C_{2212} \\
    C_{1211} & C_{1222} & C_{1212}
    \end{bmatrix} =
    \begin{bmatrix}
    \lambda + 2\mu & \lambda & 0 \\
    \lambda & \lambda + 2\mu & 0 \\
    0 & 0 & \mu
    \end{bmatrix}
\end{equation}
where $\lambda$ and $\mu$ are the Lam\'e coefficients. From these we obtain the in-plane Young's modulus and Poisson ratio~\cite{seung_defects_1988}
\begin{equation}
    Y_{2D} = \frac{4\mu\left(\mu+\lambda\right)}{2\mu + \lambda}, \hspace{4mm} \nu = \frac{\lambda}{2\mu + \lambda}.
\end{equation}
On the other hand, a discrete lattice has an energy per spring given by
\begin{equation}
    E_s\left(\mathbf{r}_{ij}\right)= \frac{1}{2} \left(\frac{1}{2} \frac{\overline{A}}{A_0} k_s\right) \left(s_{ij} - \left| \mathbf{x}_i - \mathbf{x}_j\right|\right)^2
\end{equation}
for two nodes at $\mathbf{x}_i$, $\mathbf{x}_j$ connected by a spring of length $s_{ij}$ and spring constant $k_s$. $\overline{A}$ is the sum of the facet areas adjacent to edge $\mathbf{r}_{ij}$, and $A_0$ is the area of an equilateral triangle with side length $s_{ij}$. The term $\overline{A}/(2A_0)$ is given by Van Gelder~\cite{van_Gelder_1998}, with typos in the original model corrected by Lloyd et al.~\cite{lloyd_identification_2007}.
}

\revb{When every triangle in the mesh is equilateral, i.e.\@ the nodes are arranged periodically, the energy density can be written as~\cite{ostoja-starzewski_lattice_2002}
\begin{equation}
    u_s = \frac{1}{2V} \sum_b \alpha^{(b)}\left(\ell^{(b)}\right)^2 n_i^{(b)} n_j^{(b)} n_k^{(b)} n_m^{(b)} \epsilon_{ij} \epsilon_{km}
\end{equation}
where $V$ is the volume of a cell surrounding a node, $b$ is the index of a bond connected to the node, $\alpha$ is the spring constant of a half-length bond, and $\ell$ is the rest length of a half-length bond. The vectors $n_i$ are unit vectors parallel to the bond $b$, and $i \in \{1,2\}$; $\epsilon_{ij}$ are components of the in-plane strain tensor, and Einstein summation notation is used.
Given this definition of energy density, we can identify that the stiffness tensor is
\begin{equation}
    C_{ijkm} = \frac{1}{V} \sum_b \alpha^{(b)} \left(\ell^{(b)}\right)^2 n_i^{(b)} n_j^{(b)} n_k^{(b)} n_m^{(b)}.
\end{equation}
In the regular (hexagonal) lattice $\overline{A}/(2A_0)$ reduces to unity, $\ell = s/2$, and the cell volume $V = 2 \sqrt{3} \ell^2$. All the springs have the same constant $\alpha=2k_s$. Equating the continuous and discrete forms of the stiffness tensor gives
\begin{equation}
    \lambda = \mu = \frac{\sqrt{3}}{4} k_s
\end{equation}
as obtained in previous work~\cite{seung_defects_1988,ostoja-starzewski_lattice_2002}. Thus the in-plane Young's modulus and Poisson ratio for a regular (hexagonal) latice are
\begin{equation}
    Y_{2D} = \frac{2}{\sqrt{3}}k_s, \hspace{4mm} \nu = \frac{1}{3}.
    \label{eq:hex_mod}
\end{equation}}\revb{Typically for a non-regular lattice, $\overline{A}/(2A_0) \neq 1$, so the quantities in Eq.~\eqref{eq:hex_mod} no longer hold beyond a first order approximation in the discretization process. Indeed for non-regular mesh topologies, one cannot write an exact formula for the in-plane Young's modulus. Instead, we choose a target $Y_{2D} = 2 k_s / \sqrt{3}$ for the sheet so that the stretching energy has the form
\begin{equation}
  E_s\left(\mathbf{r}_{ij}\right) = \frac{1}{2} \left(\frac{\sqrt{3}}{4} \frac{\overline{A}}{A_0} Y_{2D}\right)\left(s_{ij} - \left|\mathbf{x}_i-\mathbf{x}_j\right|\right)^2, \label{eq:mVG_app}
\end{equation}
as previously stated in Eq.~\eqref{eq:rand_stretch_eng}. We show in Appendix \ref{appendix:mod_tests} that as the mesh size decreases, the bulk in-plane modulus for the sheet converges to the target value for $Y_{2D}$, and a Poisson ratio of $\nu = 1/3$. In other words, in the absence of an exact translation between a continuum model and a sheet with randomly placed nodes, we can approximate an isometric sheet with a given modulus and $\nu = 1/3$. This is done by allocating the stiffness of each spring constant based on the local topology of the mesh, such that all calculations are local and still rely only on terms that are first order in the deformation.
}

\subsection{Bending}
\revb{Still following the SN model~\cite{seung_defects_1988}, we begin with the total bending energy for a continuous sheet with area $S$, embedded in $\mathbb{R}^3$:
\begin{equation}
    U_b = \frac{1}{2} B \int_S H^2 dA,
\end{equation}
where $B$ is the bending rigidity, and $H$ is the mean curvature (calculated as the sum of principal curvatures at a point). When the surface is deformed by a small amount $f$, the mean curvature is approximately the Laplacian: $H \approx \nabla^2 f$.
}

\revb{For a discrete mesh, we use the bending energy
\begin{equation}
    E_b\left(\hat{\mathbf{n}}_{ijk},\hat{\mathbf{n}}_{ikl}\right) = \frac{1}{2} \left( 2 \frac{A_0}{\overline{A}} k_b \right) \left|\hat{\mathbf{n}}_{ijk} - \hat{\mathbf{n}}_{ikl}\right|^2,
\end{equation}
where $\hat{\mathbf{n}}_{ijk}$ and $\hat{\mathbf{n}}_{ikl}$ are the unit normals to triangles with vertices $ijk$ and $ikl$ respectively; $\overline{A}$ is the sum of facet areas adjacent to the edge $ik$ between the two facets, and $A_0$ is the area of an equilateral triangle with side length $s_{ik}$. Essentially this bending energy is a penalization of misaligned normals for neighboring facets. The prefactor $2 A_0/\overline{A}$ is inspired by Wardetzky et al.~\cite{wardetzky_discrete_2007}, who chose it as a quantification of the shape (and thereby mass distribution) of the triangle facets adjacent to the edge in question. Although we present it as a ratio of areas, the quantity encodes the local topology of the mesh.
}

\revb{In the case of a mesh composed entirely of equilateral triangles, the prefactor $2 A_0/\overline{A}$ again reduces to unity, so the discrete bending energy is
\begin{equation}
    E_b\left(\hat{\mathbf{n}}_{ijk},\hat{\mathbf{n}}_{ikl}\right) = \frac{1}{2} \left|\hat{\mathbf{n}}_{ijk} - \hat{\mathbf{n}}_{ikl}\right|^2,
    \label{eq:SN_bending}
\end{equation}
in agreement with the SN model. By rolling a mesh of equilateral triangles into a cylinder, Seung and Nelson related the continuous bending rigidity $B$ to the discrete bending constant $k_b$~\cite{seung_defects_1988}:
\begin{equation}
    B = \frac{\sqrt{3}}{2} k_b.
\end{equation}
This relationship does not hold for non-regular mesh topologies, but as in the stretching case, we choose a target $B = \sqrt{3}/(2k_b)$ for the sheet, which gives the bending energy the form
\begin{equation}
    E_b\left(\hat{\mathbf{n}}_{ijk},\hat{\mathbf{n}}_{ikl}\right) = \frac{1}{2} \left( \frac{4}{\sqrt{3}} \frac{A_0}{\overline{A}} B \right) \left|\hat{\mathbf{n}}_{ijk} - \hat{\mathbf{n}}_{ikl}\right|^2,
    \label{eq:mG_bending}
\end{equation}
previously stated in Eq.~\eqref{eq:rand_bend_eng}, which we call the modified Grinspun (mG) model. Appendix \ref{appendix:mod_tests} demonstrates that as the mesh size tends to zero, the bulk bending modulus of the sheet converges to the target value for $B$. By allocating the stiffness of each hinge spring according to the surrounding mesh topology, we can use this local deformation-based model to approximate a continuous sheet with rigidity $B$.
}

\revb{There have been critiques of Seung and Nelson's treatment of the Gaussian rigidity in mapping the continuous bending model to the discrete one~\cite{schmidt_universal_2012}. However, in spite of this critique of the SN bending model, we find that both the SN model (Eq.~\eqref{eq:SN_bending}) and the mG model (Eq.~\eqref{eq:mG_bending}) perform well numerically and demonstrate convergence as a function of mesh size, as discussed in Appendix~\ref{appendix:mod_tests}.}

\section{Numerical details}\label{appendix:numerics}
Here, we briefly summarize the numerical approach used to model the twisting of thin ribbons in this work. Simulations of twisted ribbons are performed using a custom code developed previously to study the crumpling of thin sheets~\cite{andrejevic_simulation_2022}. The code is implemented in C++ and multithreaded using OpenMP~\cite{dagum1998openmp}. As noted in the main text, the mesh topology of the ribbons is generated by randomly seeding a rectangular domain with nodes and constructing a Delaunay triangulation via the Voro++ library~\cite{rycroft_voro_2009,lu22}. The random triangulation is regularized using Lloyd's algorithm, an iterative method that incrementally steers the triangles to a more uniform shape and size~\cite{lloyd1982}.

To evolve the dynamics of the ribbon in time under applied tension and twist, the equations of motion~(Eq.~\eqref{eq:eom}) are solved numerically using a hybrid integration scheme, as presented in Ref.~\cite{andrejevic_simulation_2022}. In this approach, we recognize that the ribbon deforms slowly and smoothly for the majority of the simulation, which permits a quasistatic approximation of the equations of motion such that $\bm{a}_i\approx \bm{0}$ for a node $i$ in the sheet. The resulting equations of motion,
\begin{align}
    \begin{split}
        \dot{\bm{x}}_i &= \bm{v}_i, \\
        \bm{F}_i &= \bm{0},
    \end{split}
\end{align}
describe a differential-algebraic system of equations, or DAE, which contains both differential equations ($\dot{\bm{x}}_i = \bm{v}_i$) and algebraic constraints ($\bm{F}_i = \bm{0}$) that must be simultaneously satisfied. Due to the presence of algebraic constraints, DAEs are typically solved using implicit methods, and a backward differentiation formula (BDF) is used to discretize and integrate this system. The solution vector $(\bm{x}_i,\bm{v}_i)$ at each new timestep is computed iteratively using Newton's method, and each Newton iteration entails a solving a linear system. The linear systems are solved using the conjugate gradient method, which is well-suited for large, sparse, symmetric, positive-definite systems as occur in our problem. Small performance boosts are also obtained through preconditioning of the linear system and are described in greater detail in Ref.~\cite{andrejevic_simulation_2022}.

While a majority of the deformations during twisting, such as the onset of longitudinal wrinkles, are smooth, a rapid deformation occurs during transverse buckling of the sheet, and possibly self-contact. In these cases, the local velocity at any point in the ribbon may be large, and the quasistatic approximation no longer holds. The breakdown of the quasistatic approximation is detected by monitoring the maximum rate of change in velocity at each timestep and identifying if it exceeds a specified threshold. When the threshold is exceeded, the fully dynamic equations of motion (Eq.~\eqref{eq:eom}) are solved instead using a standard explicit Runge--Kutta method. Both the implicit and explicit methods employ adaptive step control by measuring the discrepancy between a lower and higher order solution at each step. We refer to Ref.~\cite{andrejevic_simulation_2022} for complete details on this method.

\section{Model validation}
\label{appendix:mod_tests}

Three modulus tests are employed to probe the response of both types of spring models under different loading conditions across various mesh resolutions: uniaxial stretching; uniformly loaded, simply supported bending; and pure shearing. In all tests the sheet is loaded with an appropriate stress to produce a deformation well within the Hookean regime for the sheet (i.e.\@ well-modeled by an analytical solution linear in $\varepsilon$); in every test the
target strain is $\varepsilon = 2.5 \times 10^{-5}$. These tests probe the small-deformation stress--strain response of our sheet and demonstrate convergence as a function of mesh resolution for both spring models.

To test how well our MSM produces the expected bulk material properties, we track the displacement of each node in the mesh. Ideally the discretized mesh will deform exactly as a continuous sheet would under the specified loading conditions. We compare the actual position of each node to the final homogeneous-deformation position. The error of the entire mesh is then calculated as a scaled $L^2$ norm using the formula
\begin{equation}
    \label{eq:error_def}
    E = \sqrt{\frac{1}{3A} \sum_{i=1}^N A_i^t \lVert \bm{x}_i - \bm{x}_i^h\rVert_2^2 }
\end{equation}
where $A$ is the area of the entire sheet, the sum is over each node $i$, $A_i^t$ is the area of the undeformed triangles adjacent to a node, $\bm{x}_i$ is the position of a node, and $\bm{x}_i^h$ is the position the node should have under a homogeneous deformation. The pre-factor of $1/3$ counteracts the triple counting of $A_i^t$.

We fit the error to the form
\begin{equation}
    \label{eq:error_fit}
    E = a d^b
\end{equation}
where $d$ is the characteristic length scale of the mesh (mean spacing of the nodes, or exact spacing in the case of a regular lattice) and $a$ and $b$ are fitting parameters. The value of $b$ tells us the order of convergence of the numerical method with the chosen spring model. Two types of meshes were tested: a regularly packed triangular lattice, and a randomly seeded triangular lattice. The regular meshes were generated to have similar non-equilateral triangles at each corner, to reduce the errors contributed by non-regular spring lengths at the boundary. Five sets of random meshes with different seeds were generated to fit the convergence across the random-type mesh.

\begin{table}[h]
    \centering
    \renewcommand{\arraystretch}{1.5}
    \begin{tabular}{l|c|c}
    Mesh type \hspace{5mm} & \, Aspect Ratio \, & %\, Length \, & \, Width \, &
    \, Node spacing range (mm)\\ \hline Regular A & 0.996 & %0.0996 & 0.1 &
    [0.500, 2.50] \\ Random A & 1.00 & %0.1 & 0.1 &
    [0.577, 2.24]\\ Regular B & 8.66 & %0.0866 & 0.01 &
    [0.600, 5.00]\\ Random B & 10.0 & %0.1 & 0.01 &
    [0.707, 3.16]
    \end{tabular}
    \caption{Dimensions of the four test meshes used to perform the three types of modulus tests. All meshes had Young's modulus $Y = \SI{1.00}{\giga \pascal}$ and thickness $h = \SI{1.00}{\milli \meter}$. The theoretically predicted value of Poissson ratio for any triangular lattice is $\nu = 1/3$~\cite{seung_defects_1988,lloyd_identification_2007}. Square test meshes (type A) were used for the stretching and shearing tests, whereas long, thin test meshes (type B) were used in the bending tests.}
    \label{tab:meshes}
\end{table}

Table~\ref{tab:meshes} gives the dimensions of the test meshes used. ``A'' type meshes were used for the stretch and shear tests, ``B'' types were used in the bend test.
All sheets had a Young's modulus of $Y = \SI{1.00}{\giga \pascal}$ and thickness of $h = \SI{1.00}{\milli \meter}$ (with corresponding bending rigidty $B = \SI{3.00}{\newton \cdot \meter}$).

\begin{figure}
    \centering
    \includegraphics[width=8.6cm]{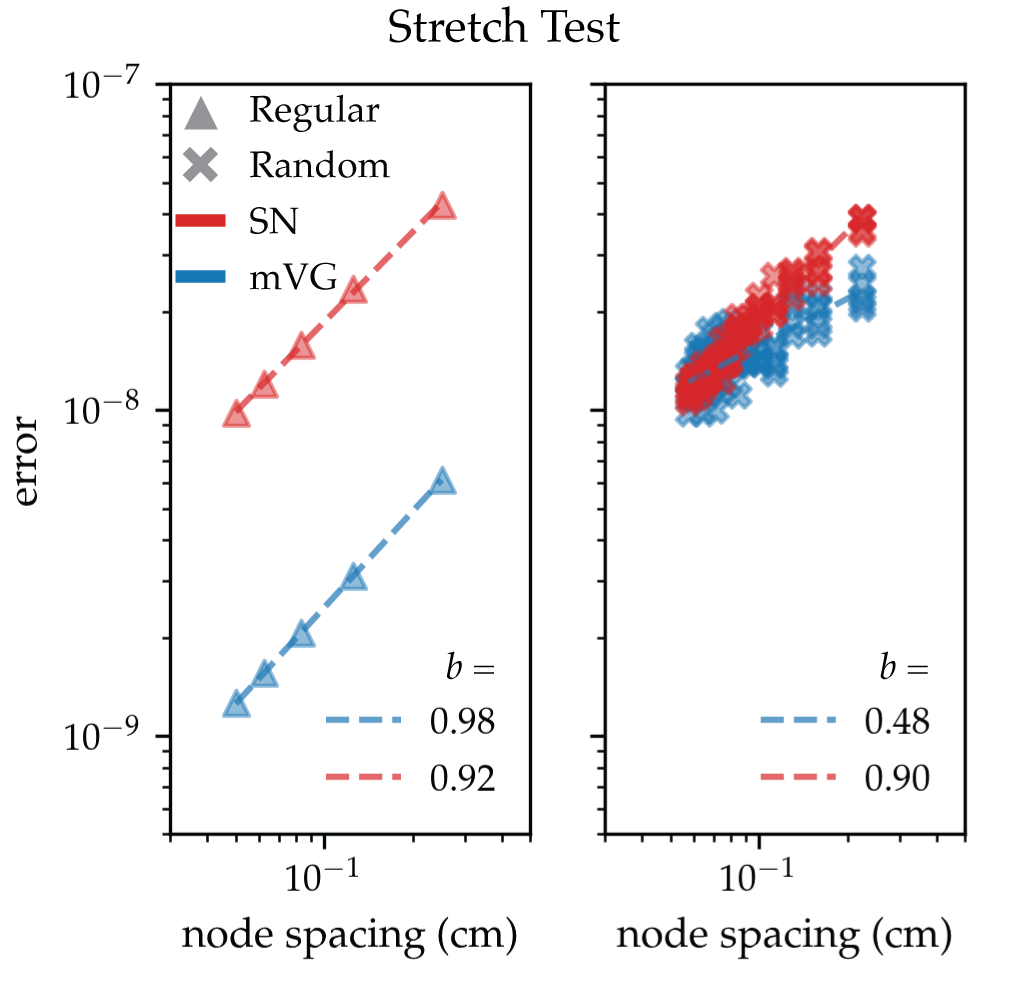}
    \caption{Uniaxial stretching errors in the regular (left, triangular markers) and random (right, X markers) meshes as a function of mesh resolution. For both meshes the mVG model (blue) error has a smaller magnitude than the SN model (red).}
    \label{fig:stretch_test}
\end{figure}

\begin{figure}
    \centering
    \includegraphics[width=8.6cm]{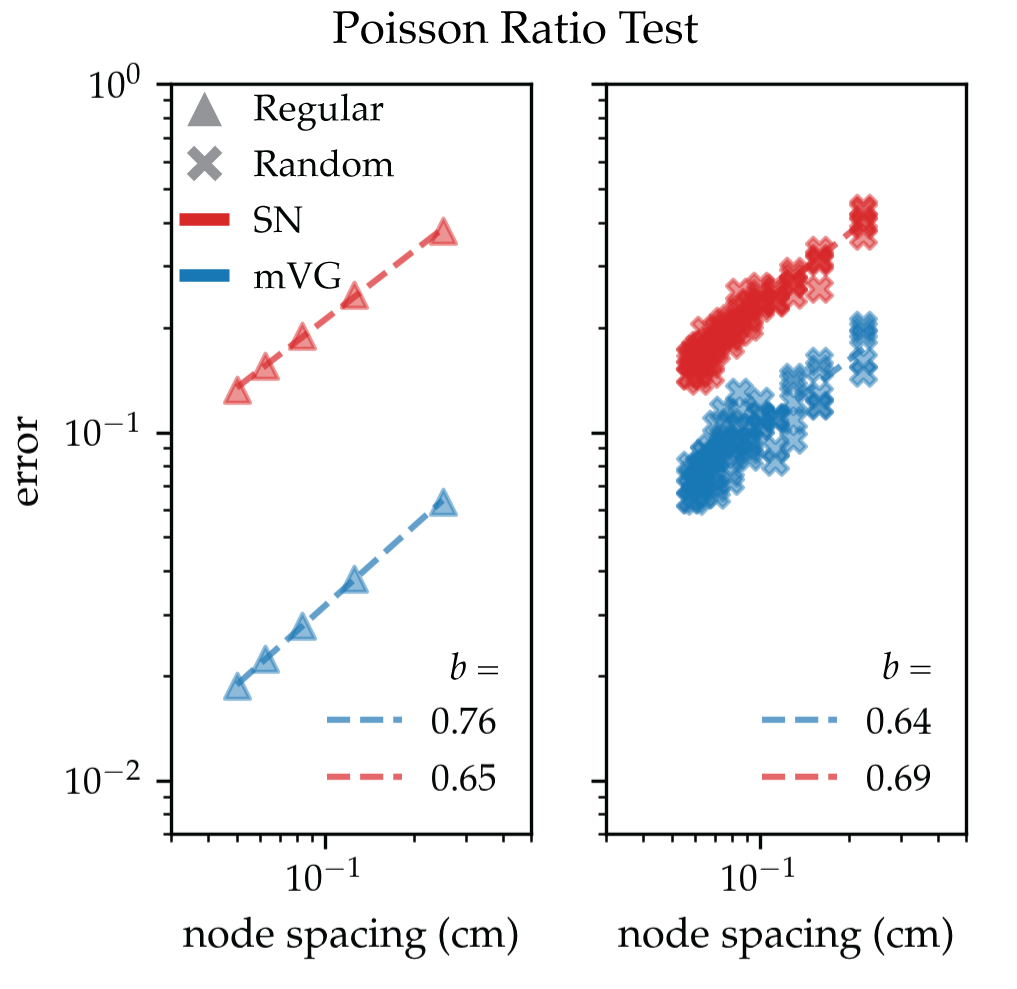}
    \caption{Percent errors in the measured Poisson ratio compared to the expected value of $\nu = 1/3$ in the regular (left, triangular markers) and random (right, cross markers) meshes. The mVG model (blue) more reliably produces the expected Poisson ratio than the SN model (red).}
    \label{fig:poisson_ratio_test}
\end{figure}

\begin{figure}
    \centering
    \includegraphics[width=8.6cm]{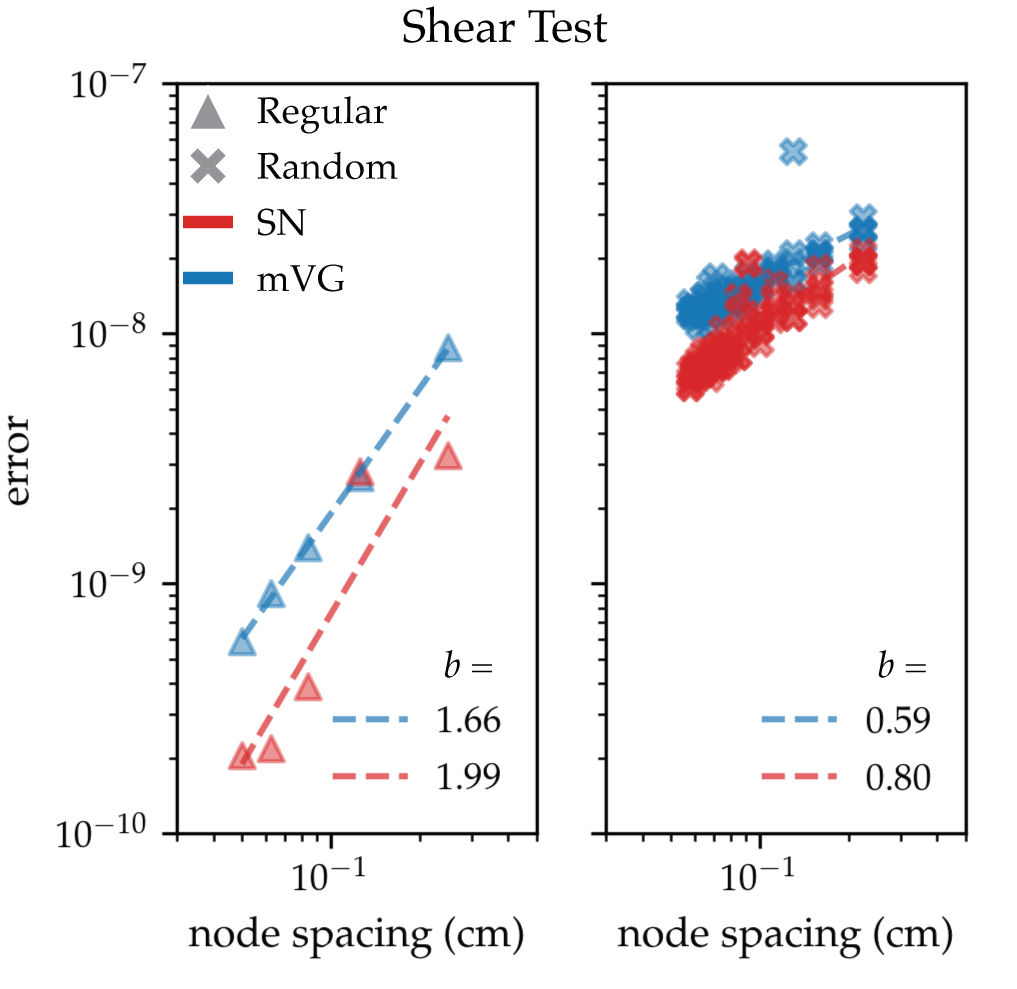}
    \caption{Pure shearing errors in the regular (left, triangular markers) and random (right, cross markers) meshes as a function of mesh resolution. For both meshes the SN model (red) has smaller magnitude of errors and converges at a higher order than the mVG model (blue).}
    \label{fig:shear_test}
\end{figure}

The stretching and shearing effects are dependent only on the sheets' in-plane springs. In Figs.~\ref{fig:stretch_test}--\ref{fig:shear_test}, we test the SN model for stretching (Eq.~\eqref{eq:stretch_eng}) versus the modified Van Gelder (mVG) model (Eqs.~\eqref{eq:rand_stretch_eng} \& \eqref{eq:mVG_app}).

The expected deformation of the sheet under uniaxial stretching in the $y$ direction is
\begin{equation}
    \label{eq:stretch}
    F_\text{stretch}\left(x_0,y_0,z_0\right) \to \left( \left(1- \nu \varepsilon_\text{st}\right)x_0, \left(1+\varepsilon_\text{st}\right)y_0, z_0\right)
\end{equation}
where $\varepsilon_\text{st} = \frac{\Delta L}{L}$ is the target strain. The stress applied to the edges of the sheet is thus $\sigma_\text{st} = Y\varepsilon_\text{st}$. Figure \ref{fig:stretch_test} plots the error, Eq.~\eqref{eq:error_def}, across a range of mesh resolutions. On the left, in triangular markers, are the errors for a regular mesh, and on the right, with cross markers, are the random mesh errors across all five random sheets. For both mesh topologies the mVG model has smaller magnitude of error than the SN model, and for the regular mesh, the mVG model also converges at a slightly greater order.

As an additional metric, during the stretch test we also measure the percent error in the actual Poisson ratio compared to the analytically dictated value of $\nu = 1/3$~\cite{seung_defects_1988,lloyd_identification_2007}. Here the percent error of $\nu$ is calculated from bulk measurements of the sheets' width and length. Figure \ref{fig:poisson_ratio_test} shows that in all cases the mVG model more consistently reproduces a mesh with $\nu = 1/3$.

A sheet subjected to pure shear along all four edges should deform as
\begin{equation}
    \label{eq:shear}
    F_\text{pure shear} \left(x_0,y_0,z_0\right) \to \left( x_0 + \varepsilon_\text{sh} y_0, \varepsilon_\text{sh} x_0 + y_0, z_0\right)
\end{equation}
where $\varepsilon_\text{sh} = \frac{1}{2}\left(\frac{\Delta W}{L} + \frac{\Delta L}{W}\right)$ is the target strain. The shear stress applied to the edges is therefore $\sigma_\text{sh} = 2 G \varepsilon_\text{sh}$, with $G$ the shear modulus. In this case we see in Fig.~\ref{fig:shear_test} that the SN model has errors of smaller magnitude than the mVG model and also converges at a greater order. However, in the regular mesh case (left, triangular markers), the mVG model performs more consistently than the SN model.

Throughout this paper we implemented the mVG model for stretching because of its advantages in reproducing the expected Young's modulus and Poisson effect. Although the mVG model has a disadvantage under shear deformations, shear is not particularly relevant for our twisted ribbon studies. Further, the differences between the models are slight for the random meshes, so we concede one less-relevant disadvantage to employ two salient advantages.

\begin{figure}
    \centering
    \includegraphics[width=8.6cm]{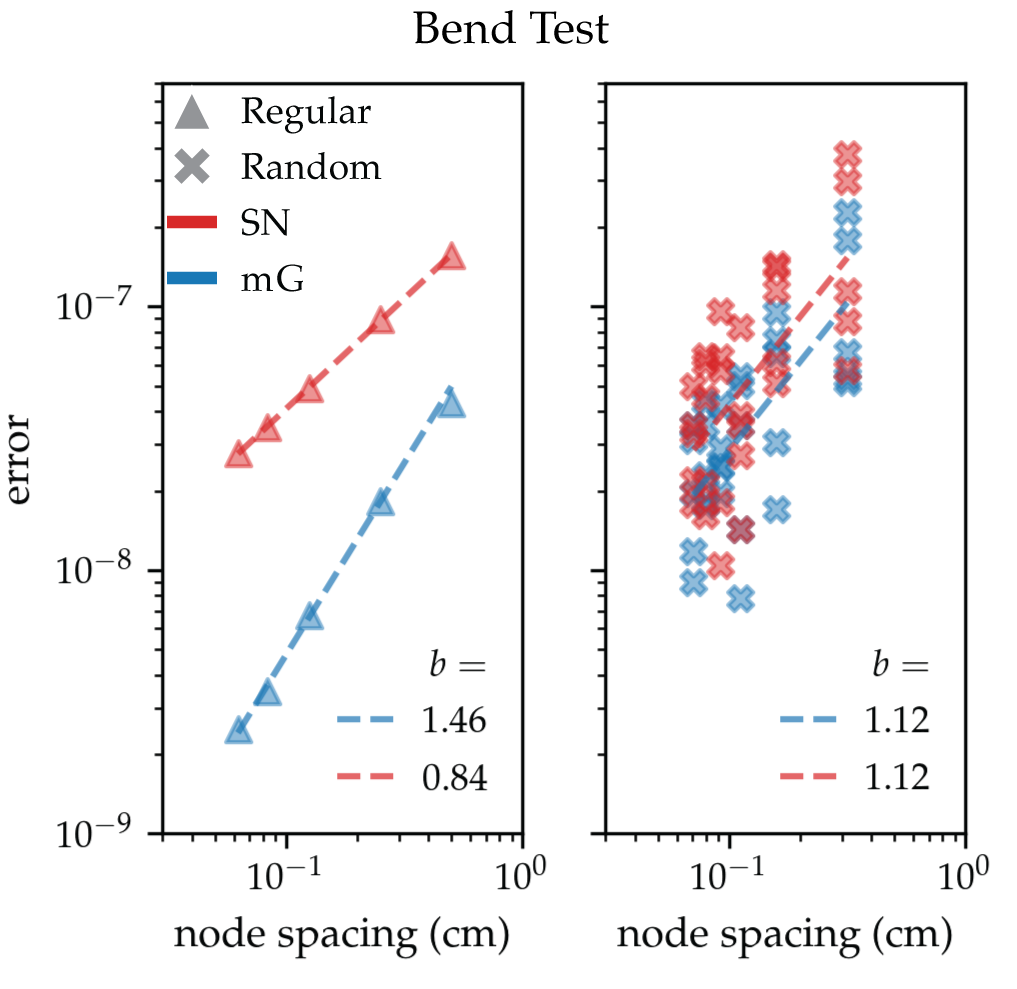}
    \caption{Uniformly loaded, simply supported plate bending errors in the regular (left, triangular markers) and random (right, X markers) meshes as a function of mesh resolution. For the regular mesh the mG model (blue) has a smaller magnitude of error and greater order of convergence than the SN model (red). The random meshes converge at the same order, but in general the magnitude of the error is smaller when using the mG model.}
    \label{fig:bend_test}
\end{figure}

The bending effects are dependent only on the sheets' out-of-plane rigidity, from the hinge edge pseudo-springs. As shown in Fig.~\ref{fig:bend_test}, we test the SN model for bending (Eq.~\eqref{eq:bend_eng}) versus the mG model (Eqs.~\eqref{eq:rand_bend_eng} \& \eqref{eq:mG_bending}).

A uniformly loaded, simply supported plate (rotation is allowed at the free edges, but edges are fixed in the $z$ direction) has a bent profile of
\begin{equation}
    \label{eq:bend_profile}
    z_\text{ss}\left(y\right)= - \frac{\varepsilon_\text{b}}{5 L^3} \left(5 L^4 - 24 L^2 y^2 + 16 y^4\right)
\end{equation}
where $y \in [-L/2,L/2]$ and $\varepsilon_\text{b} = \frac{\Delta z_\text{max}}{L}$ is the target strain. The stress applied across the sheet is $\sigma_\text{b} = \frac{1024}{15} \frac{B}{L^3}\varepsilon_\text{b}$, with $B$ the bending rigidity. Therefore the expected deformation of the sheet is
\begin{equation}
    \label{eq:bend}
    F_\text{ss bend}\left(x_0,y_0,z_0\right) \to \left( x_0, y_0, z_\text{ss}\left(y_0\right)\right)\, .
\end{equation} ~\\
Figure \ref{fig:bend_test} shows that the mG model performs better in the regular mesh than the SN model. For random meshes the models converge at the same order, but generally the mG error has a smaller magnitude. Thus we implement the mG model for bending in this paper because its error is generally less than the SN bending model.

While we find the definition of error in Eq.~\eqref{eq:error_def} to most thoroughly quantify the amount of error across the sheet, we can also calculate the percent error of the modulus of interest. The average node spacing of our ribbon meshes is $d = \SI{1}{\milli \meter}$. For a test mesh of similar node spacing, the mVG/mG models give average error $\approx 4.5\%$ for Young's modulus, $\approx 1.5\%$ for shear modulus, and $\approx 1.9\%$ for bending modulus.
\end{document}